\documentclass{article}
\usepackage[utf8]{inputenc}
\usepackage{cite}
\usepackage{graphicx}
\usepackage{subfigure}
\usepackage{algorithm}
\usepackage{algpseudocode}
\usepackage{appendix}
\usepackage{amsmath}
\usepackage{authblk}
\usepackage{soul}
\usepackage{indentfirst}
\usepackage[normalem]{ulem} 
\usepackage[dvipsnames,table]{xcolor}
\usepackage{float}
\usepackage{caption} 
\usepackage[table]{xcolor}
\graphicspath{ {./images/} }
\providecommand{\keywords}[1]{\textbf{\textit{Keywords---}} #1}

\begin{document}
\title{Stochastic Nonlinear Model Updating in Structural Dynamics Using a Novel Likelihood Function within the Bayesian-MCMC Framework}

\author[1]{Pushpa Pandey}
\author[1*]{Hamed Haddad Khodaparast}
\author[1]{Michael Ian Friswell}
\author[2]{Tanmoy Chatterjee }
\author[1]{Hadi Madinei}
\author[3]{Tom Deighan}
\affil[1]{\normalsize Faculty of Science and Engineering, Swansea University, Wales, United Kingdom}
\affil[2]{\normalsize School of Mechanical Engineering Sciences, University of Surrey, England, United Kingdom}
\affil[3]{\normalsize United Kingdom Atomic Energy Authority, United Kingdom}
\affil[1*]{\normalsize Corresponding Author:Hamed Haddad Khodaparast, h.haddadkhodaparast@swansea.ac.uk}
\date{April 2024}

\maketitle

\begin{abstract}
The study presents a novel approach for stochastic nonlinear model updating in structural dynamics, employing a Bayesian framework integrated with Markov Chain Monte Carlo (MCMC) sampling for parameter estimation by using an approximated likelihood function. The proposed methodology is applied to both numerical and experimental cases. The paper commences by introducing Bayesian inference and its constituents: the likelihood function, prior distribution, and posterior distribution. The resonant decay method is employed to extract backbone curves, which capture the non-linear behaviour of the system. A mathematical model based on a single degree of freedom (SDOF) system is formulated, and backbone curves are obtained from time response data. Subsequently, MCMC sampling is employed to estimate the parameters using both numerical and experimental data. The obtained results demonstrate the convergence of the Markov chain, present parameter trace plots, and provide estimates of posterior distributions of updated parameters along with their uncertainties. Experimental validation is performed on a cantilever beam system equipped with permanent magnets and electromagnets. The proposed methodology demonstrates promising results in estimating parameters of stochastic non-linear dynamical systems. Through the use of the proposed likelihood functions using backbone curves, the probability distributions of both linear and non-linear parameters are simultaneously identified. Based on this view, the necessity to segregate stochastic linear and non-linear model updating is eliminated.

\end{abstract}

\section{Introduction}
The deterministic linear model updating \cite{MARES20061674, Friswell1995} process in structural dynamics has become an established procedure in the industry. The model updating problem was initially introduced for deterministic and linear models. Its objective was to fine-tune numerical or analytical model parameters to enhance the correlation between the model’s predictions and the experimentally observed behaviour of the system. In the context of linear deterministic model updating, the process involves using dynamic data (such as modal parameters) to adjust parameters in a linear model, thereby improving correlation with test data. This typically employs sensitivity-based methods such as optimization algorithms to minimize discrepancies between analytical and experimental data \cite{MOTTERSHEAD20112275}. In deterministic model updating methods, the variability in the measured data is assumed to be epistemic uncertainty, primarily attributed to measurement errors, such as measurement noise. Statistical methods, for example, the minimum variance method \cite{collins1974, FRISWELL1989143}, are employed to identify the updating parameters.\

However, variability due to the assembly process, such as the production of multiple cars built identically with the same design parameters, is considered aleatory. Therefore, variations in experimental data must be accurately represented in the numerical model. To address this, stochastic linear model updating methods \cite{FONSECA2005587, Hua2008, Khodaparast2008, Khodaparast2011}, have been developed to determine the physical distribution of the updating parameters that represent variability in measurement data.  Over the last decade, there has been a significant number of publications focused on advancing stochastic linear model updating methods, as exemplified by various papers, including \cite{Sai2009, BANSAL2017481, JACQUELIN2012262, ZHOU2021, Bi2022, ROCCHETTA2018, Ming2021, ZHENG2022, SCHNEIDER2022, Zhang2023}. 
Stochastic linear model updating aims to improve the accuracy of uncertain parameters in a linear model by fusing experimental measurements with prior knowledge using probability theory \cite{SIMOEN2015123, Govers2010}. It considers the innate variability in the system, tests, and model \cite{Goulet2013}. Uncertainty is represented by modelling parameters as random variables with probability distributions. Common techniques include Bayesian inference \cite{Beck1998, Beck2002}, Gaussian processes \cite{Yuen2010}, and polynomial chaos expansions \cite{ni2019using}. Measurement noise and variability are also modelled probabilistically \cite{SIMOEN2015123}.

In previous stochastic model updating methods, the assumption of a linear model was made. However, engineering structures are often subjected to high dynamic loads, and multiple units are frequently manufactured. The primary motivation of this work is to extend the concept of stochastic linear model updating to what the authors refer to as stochastic non-linear model updating in structural dynamics. In non-linear dynamic systems, a significant challenge is the dependency of the dynamic response on the excitation force. To address this, the paper introduces the novel concept of utilizing the backbone curve within the Bayesian framework for this purpose.\

In recent years, research in non-linear model updating has expanded, with greater attention placed on deterministic approaches for this task. Researchers have increasingly studied deterministic non-linear model updating techniques utilizing non-linear structural models of dynamic systems and updating their parameters to match experimental vibration data. Concentrated efforts were made to leverage deterministic optimization to tune non-linear parameters to minimize the error between analytical and physical measurements of structural response. In the last few years, substantial progress has been made in deterministic non-linear model updating for vibration-based identification and damage assessment of structurally dynamic components and assemblies across various disciplines \cite{Borgi2008, Marwala2010, Behmanesh2014, ASGARIEH2014, Zhang2023 }.\

While deterministic non-linear model updating techniques have made substantial progress, relying solely on these methods has its constraints. Real-world structural systems display dynamic response randomness and uncertainty due to factors like complex loads, component degradation, and environmental fluctuations. Recent works on stochastic non-linear model updating have become essential for capturing this behaviour, addressing the limitations of deterministic methods \cite{Ludovic2018, mcgurk2023, DING2022, XIN2019}. This integration of stochastic approaches into non-linear models enables researchers to account for statistical variations in structural responses, enhancing robustness and predictability.\

At its core, stochastic model updating involves identifying the probabilistic properties of parameters within computational models. This transforms the updating process into one of stochastic parameter identification based on measured dynamic response data. Rather than optimizing parameters to precisely match experimental vibrations, stochastic identification aims to tune parameter distributions and uncertainties to encapsulate the statistical variations observed physically.\

Stochastic identification of parameters in non-linear systems, referred to here as stochastic non-linear model updating, has gained significant research attention over the past decade \cite{Janjanam2022, Oliva2017, Xu2017, Su2016, Janot2016, Zhang2018}. Although much work has been done on stochastic model updating for linear systems, there appears to be a noticeable gap when contemplating the use of stochastic model updating for non-linear systems. The work in this paper intends to bridge that gap while creating a toolset for identifying the parameters and their variation range observed from the measured data. The methodology developed in this work for stochastic model updating can be applied to any system whether linear or non-linear provided their measurement data and computational model are available.\

Recent studies have employed Bayesian MCMC algorithms for model updating in diverse non-linear dynamic systems, demonstrating their versatility and effectiveness in handling measurement noise and model uncertainty \cite{Abhinav2018, Linden2022, Springer2019, Huang2020}. \textcolor{ForestGreen}{Zhou and Tang \cite{Review3} developed a probabilistic framework combining two-level Gaussian processes and Bayesian inference in structural dynamics for uncertainty quantification. Later they proposed a probabilistic finite element model updating framework using Bayesian inference, MCMC sampling, and Gaussian Process emulation to handle uncertainties with incomplete modal information \cite{Review1}. Yang et al. \cite{Review2} developed a two-phase adaptive MCMC method to solve Bayesian model updating problems for complex dynamic systems.} Traditional likelihood functions in the Bayesian approach often assume Gaussian error distributions, which may not be appropriate for systems with non-Gaussian noise or intrinsic variability.\

This paper introduces a novel, generalized framework for stochastic model updating in dynamic systems, addressing both linear and non-linear behaviours. Leveraging backbone curves within a Bayesian context, our approach handles non-Gaussian and complex distributions prevalent in real-world systems, offering broad adaptability across various stochastic dynamics. The method's flexibility stems from its use of data-driven backbone curves and the likelihood function derived from these curves, making it applicable to any system with empirical response data and a theoretical model. Our framework employs MCMC techniques, particularly the Metropolis-Hastings algorithm, enhancing its capability to explore complex distributions and generalize beyond traditional Gaussian assumptions. We validate this universally applicable methodology through multiple case studies, demonstrating its effectiveness in capturing parameter variability and developing predictive models that reflect observed dynamics. While MCMC methods have been extensively developed for linear systems and specific non-linear applications, our work identifies and addresses a gap for general non-linear systems. We discuss the integration of MCMC algorithms not for their advancement per se but for how they serve the specific needs of our stochastic updating framework.
This paper offers an enhanced stochastic parameter identification method applicable to both linear and non-linear systems to quantify parameter variability from measurements. Our research fills a crucial gap in stochastic modelling for systems with complex non-linear behaviour, broadening the scope of uncertainty quantification in real-world structures.\

The structure of the paper is as follows: Section 2 provides an overview of Bayesian inference, covering the likelihood function, prior distribution, and posterior distribution. It also introduces backbone curves and different methods to extract them. In Section 3, the focus is on the numerical model (single degree of freedom (SDOF) system) used. It explains the extraction of backbone curves from time response data of the numerical models and presents the results obtained from MCMC sampling. Section 4 focuses on experimental validation of the proposed methodology. It discusses the experimental setup, which involves a cantilever beam system with permanent magnets at the tips and electromagnets. Details of the implementation of MCMC sampling using the experimental data and the results obtained are also presented in this section, Finally, the paper concludes in Section 5.

\section{Theory}
\subsection{Bayesian inference}
Bayesian inference is a statistical framework that uses Bayes' Rule to update our belief about the distribution of parameters of a system based on observed data. The basic idea of Bayesian inference is to start with a prior distribution that represents our initial beliefs about the parameters of a system, and then use Bayes' Rule to update these beliefs in light of observed data. The Bayesian framework consists of three components:
\begin{itemize}
    \item The likelihood function represents the probability of observing the data given a particular set of parameter values.
    \item The prior distribution represents our initial beliefs about the parameters of the system before any data have been observed.
    \item The posterior distribution is the updated distribution that results from combining the prior distribution and the likelihood function in accordance with Bayes' Rule.
\end{itemize}

The posterior distribution can be expressed mathematically as:
\begin{equation}
    P(\boldsymbol{ \Theta} | D, M) = P(D | \boldsymbol{ \Theta}, M)\cdot P(\boldsymbol{ \Theta} | M) / P(D | M)
\end{equation}
where $\boldsymbol{ \Theta}$ is the parameter vector to be estimated, \textit{D} is the observed data, and \textit{M} is the system model. $P(\boldsymbol{ \Theta} | D, M)$ is the posterior distribution of the parameters given the data and the model, $P(D | \boldsymbol{ \Theta}, M)$ is the likelihood function, $P(\boldsymbol{ \Theta} | M)$ is the prior distribution and $P(D | M)$ is the normalization constant of the posterior distribution.\

In practice, Bayesian inference involves computing the posterior distribution for a set of candidate parameter values and then using this distribution to make inferences about the most likely values of the parameters. For the prior distribution $P(\boldsymbol{ \Theta} | M)$, the uniform distribution is assumed for each system parameter $(\boldsymbol{ \Theta})$ with their corresponding lower and upper bounds. The posterior calculation is basically the product of likelihood and prior for which backbone curves are used as measurement data.\

The novelty of this work primarily lies in the introduction of a new likelihood function for simultaneously updating both linear and non-linear parameters of a non-linear dynamical system. Standard Bayesian inference is employed to calculate the distribution of the system parameters based on the collected observations, although more advanced methods may be explored in future studies.

\begin{figure}[H]
\begin{center}
\includegraphics[scale= 0.5]{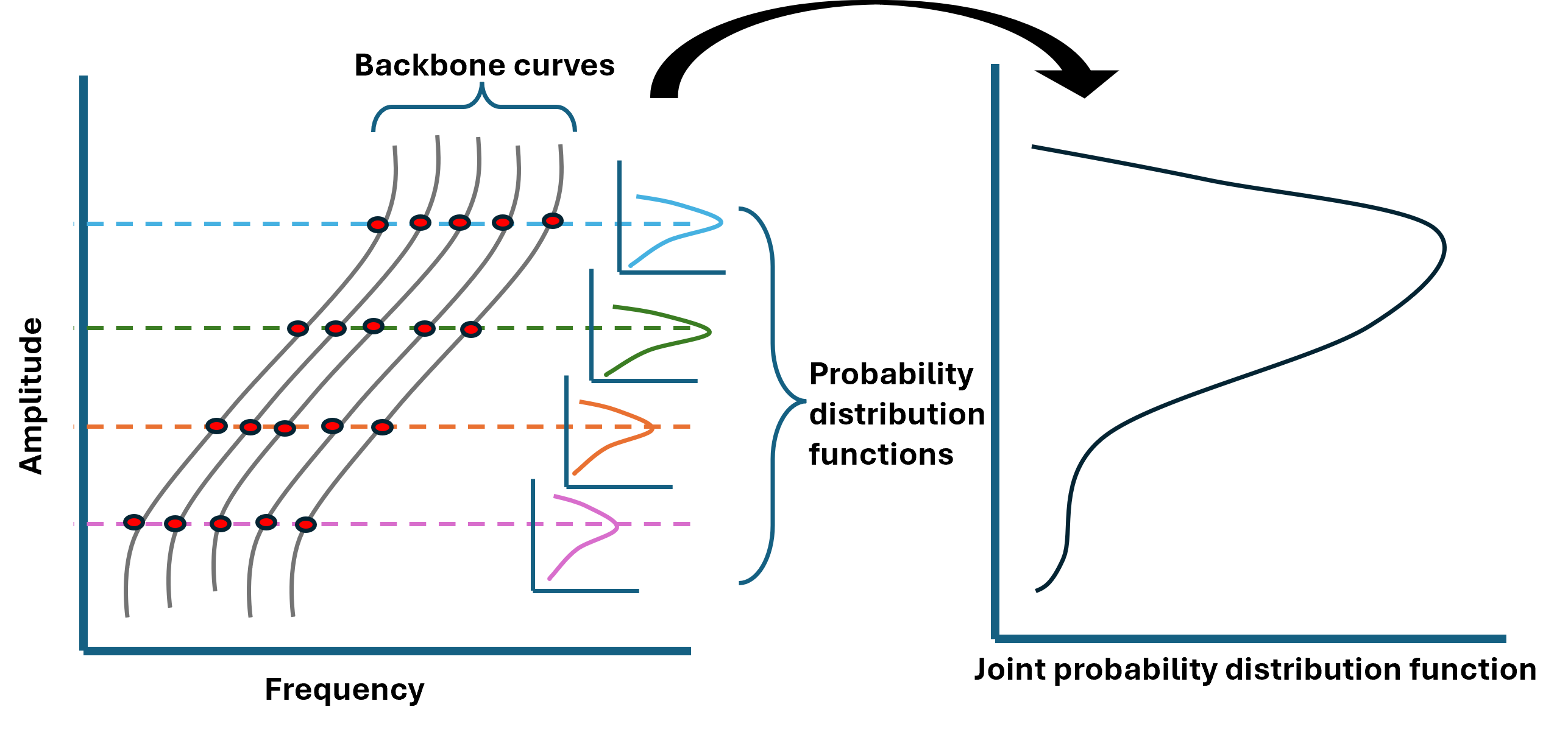}
\caption{Estimating likelihood function using a set of backbone curves}
\label{likelihood}
\end{center}
\end{figure}

\textcolor{ForestGreen}{The likelihood function was established using the backbone curves as measurement data. The left plot in Figure \ref{likelihood} shows a set of backbone curves, with red dots corresponding to points on the backbone curves at any specific amplitude. A probability density function (pdf) is fitted to all such points (red dots) lying on the set of backbone curves at an amplitude (shown by blue, green, orange, and pink pdfs). Finally, a joint pdf is taken of all such pdfs at different amplitudes, as shown in the right plot in Figure \ref{likelihood}. This joint pdf serves as the likelihood function.} 

\textcolor{ForestGreen}{Let $\boldsymbol{D} = \{d_1, d_2, \dots, d_m\}$ represent the set of measured data points along the backbone curves at different amplitudes. Each amplitude corresponds to a distribution of points, represented by red dots in Figure \ref{likelihood}. The likelihood function, $L(\boldsymbol{D} \mid \boldsymbol{ \Theta}, M)$, is defined as the joint distribution of all these distributions at different amplitudes and is given as:}

\[
L(\boldsymbol{D} \mid \boldsymbol{ \Theta}, M) = \prod_{j=1}^{N} P(d_j \mid \boldsymbol{ \Theta}, M)
\]

\textcolor{ForestGreen}{where $P(d_j \mid \boldsymbol{ \Theta}, M)$ represents the probability density function of the measured data point $d_j$, given the deterministic model and the parameter values $\boldsymbol{ \Theta}$.}

\textcolor{ForestGreen}{The multiplication of likelihood functions at different amplitudes to form the joint pdf is based on the assumption of independence between these likelihood functions. This assumption is reasonable when the selected amplitude levels are sufficiently spaced apart, encompassing distinct regimes of the system's behaviour. For instance, by choosing points from low amplitudes (where linear behaviour dominates), mid-range amplitudes (where nonlinear effects begin to emerge), and high amplitudes (where the nonlinear characteristics of the system are distinctly observable), it can be said that the system's response at these different amplitude levels is largely uncorrelated. This spacing ensures that each likelihood function captures unique information about the system's behaviour in different operational regimes, minimizing overlap and potential correlations. It allows for a comprehensive likelihood that incorporates the system’s behaviour across all amplitude ranges.}

\subsection{Markov Chain Monte Carlo (MCMC)}
\begin{figure}[H]
\begin{center}
\includegraphics[scale= 0.7]{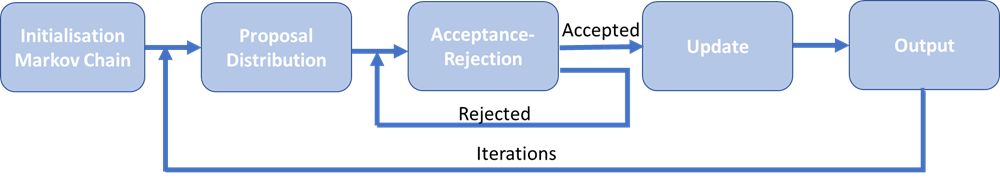}
\caption{MCMC framework}
\label{MCMCFramework}
\end{center}
\end{figure}

MCMC is a framework for generating a sequence of samples from a complex probability distribution. The samples generated by MCMC methods are used to approximate the target distribution. MCMC algorithms are particularly useful for high-dimensional distributions that are difficult to sample from directly. The MCMC framework is shown in Figure \ref{MCMCFramework}.
From Figure \ref{MCMCFramework},
\begin{itemize}
    \item Initialization represents the starting point of the algorithm, where an initial value for the Markov chain is chosen.
    \item Proposal Distribution denotes the distribution used to generate candidate states for the Markov chain. The most common proposal distribution used in the Metropolis-Hastings algorithm is the normal distribution.
    \item Acceptance-Rejection signifies the step where the candidate state is accepted or rejected based on the acceptance probability. The acceptance probability is calculated using the current state of the Markov chain, the candidate state and the proposal distribution.
    \item Update indicates the step where the current state of the Markov chain is updated to the candidate state if it is accepted. If the candidate state is rejected, the current state remains unchanged.
    \item Iteration refers to the repeated execution of the above steps until a desired number of samples are generated.
    \item Output represents the output of the algorithm, which is a set of samples from the target distribution.
\end{itemize}

\begin{algorithm}
  \caption{MCMC Sampling with Metropolis-Hastings}
    \label{alg:algo}
  \begin{algorithmic}
    \Procedure{MCMC-MH}{$\theta_0$, $T$}
      \State \textbf{Input:}
      \State \quad Initialize $\theta_0$ to a starting value
      \State \quad Set the number of iterations, $T$
      \State \textbf{Output:}
      \State \quad Samples: $\theta_1, \theta_2, \ldots, \theta_T$ from the target distribution
      \For{$t = 1$ to $T$}
        \State Sample a proposal $\theta'$ from a proposal distribution $q(\theta' | \theta_{t-1})$
        \State Calculate the acceptance ratio $\alpha$ using:
        \State \quad $\alpha = \min\left(1, \frac{p(\theta') q(\theta_{t-1} | \theta')}{p(\theta_{t-1}) q(\theta' | \theta_{t-1})}\right)$
        \State Generate a random number $u$ from a uniform distribution $U(0, 1)$
        \If{$u \leq \alpha$}
          \State $\theta_t = \theta'$ \Comment{Accept the proposal}
        \Else
          \State $\theta_t = \theta_{t-1}$ \Comment{Reject the proposal}
        \EndIf
      \EndFor
      \State \textbf{Output:} $\theta_1, \theta_2, \ldots, \theta_T$
    \EndProcedure
  \end{algorithmic}
\end{algorithm}

Except for relatively simple probabilistic models, the direct calculation of the desired parameters from any practical model is difficult to solve. This poses the need for approximating the expected probability. The solution can be either the sum of a discrete distribution or the integral of a continuous distribution with many variables, both of which are impossible to solve \cite{Bishop2006}. \

This is where Monte Carlo sampling helps in getting the solution by repeatedly drawing independent samples from the probability distribution. However, the incorrect assumptions of Monte Carlo sampling that each sample was drawn from the target distribution are independent and the method's inability to perform well with an increased number of parameters hinders the objective of this sampling method to be used for high dimensional cases.\

Combining Monte Carlo sampling with Markov Chain where the Markov chain ensures that the current value of the variable is only dependent on its prior value, random sampling of high dimensional data can be achieved. MCMC uses a sampler in the form of the Metropolis-Hastings algorithm which decides the acceptance of the samples into the chain by employing the proposal distribution. The proposal distribution is defined to evaluate and sample from. It is a conditional probability distribution whose functional form changes depending on the sampler's current state.\

The methodology proposed here using MCMC (Algorithm~\ref{alg:algo}) develops a stochastic model for the system parameters by making use of the measurement data and a computational model. The measurement data helps in establishing the likelihood function of the Bayesian approach considered here.

\subsection{Combined Bayesian and MCMC framework}
Traditional Bayesian methods struggle with non-linear stochastic dynamic systems due to complex likelihood functions, sensitivity to initial conditions, model misspecification and data limitations, which can lead to biases.
The proposed method combines Metropolis-Hastings MCMC with a likelihood function approximation using backbone curves to simplify non-linear dynamics.
This approach is poised to mitigate several existing drawbacks. By simplifying the representation of the system's dynamics through backbone curves, the complexity of the likelihood function may be significantly reduced. In terms of stability, a robustly captured backbone curve may lessen the initial condition sensitivity, leading to a more consistent likelihood function. Furthermore, this data-driven approximation might enhance the robustness against model misspecification by relying less on theoretical assumptions and more on the observed behaviour of the system. Lastly, the efficiency of data usage could be improved, which is particularly advantageous in scenarios where data is scarce. \

The methodology presented does not introduce novel features when compared with existing frameworks. It develops a stochastic model for system parameters utilizing measurement data and a computational model, where the data is employed to define the likelihood function for the Bayesian approach in use. The standard elements of the Bayesian framework—likelihood function, prior, and posterior distributions—are outlined, with MCMC techniques applied to generate samples from the posterior distribution for parameter estimation (shown in Figure 3). This conventional combination of Bayesian inference with MCMC is employed particularly where models are complex or non-linear, and where traditional optimization might not effectively reach a global optimum. The MCMC algorithm operational within this framework produces a Markov chain of samples from the posterior distribution, which is a traditional approach to quantify the uncertainty of estimated parameters, representing them as distributions. However, it must be stressed that this framework does not extend beyond the established methodologies and their inherent capabilities.

\begin{figure}[H]
\begin{center}
\includegraphics[scale= 0.4]{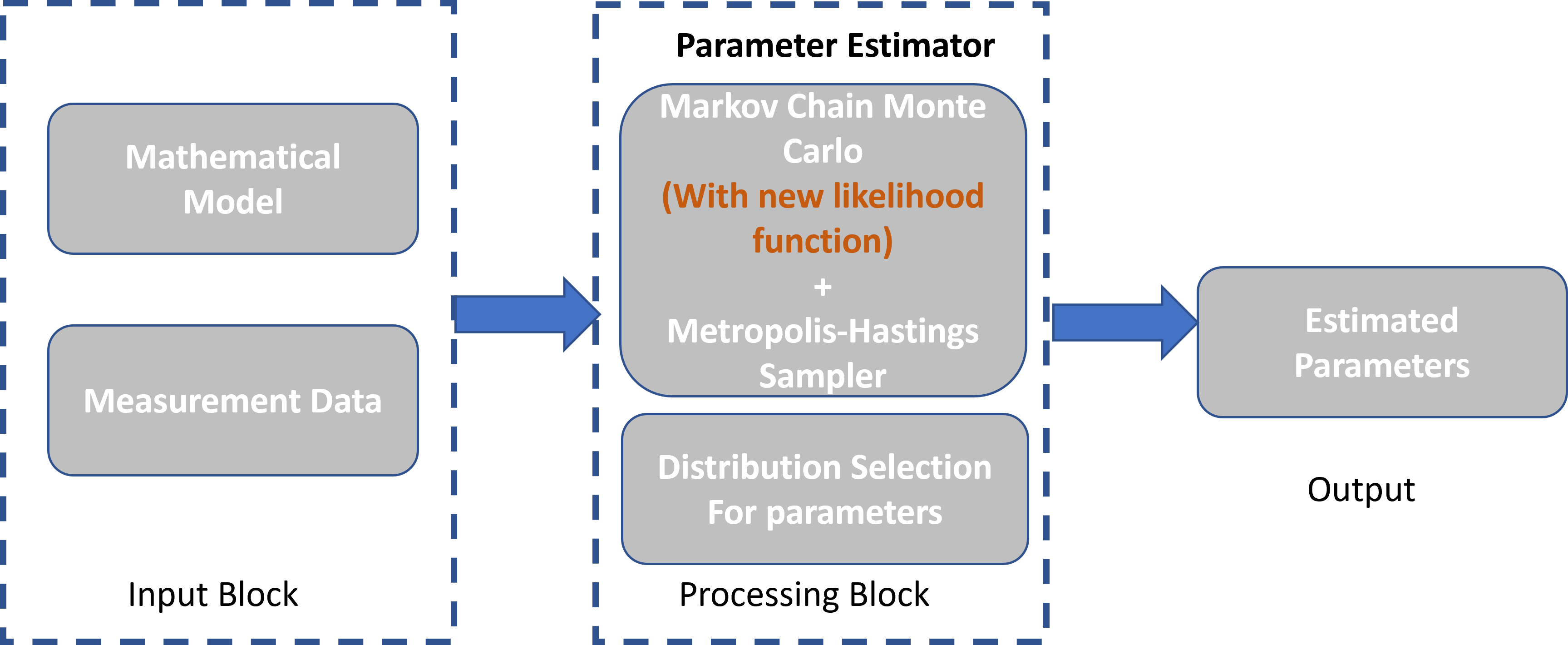}
\caption{Stochastic parameter identification model}
\label{Stochastic}
\end{center}
\end{figure}

\subsection{Backbone curves}
Backbone curves make use of the response of the non-linear system in the absence of force and damping. It reflects some of the important characteristics of the non-linearity present in the system such as modal interaction, exchange of modal energy, and non-linear model parameters. Various techniques were proposed and have been used to generate backbone curves from experimentation such as Response-Controlled stepped-sine Testing (RCT) \cite{Karaaal2020}, where the modal amplitude is kept constant and the excitation is varied, Control based continuation (CBC) \cite{Barton2015}\cite{Renson2016}, where the frequency is kept constant and the excitation level is varied, the Phase-locked-loop (PLL) control algorithm \cite{Peter2016}, where the phase difference between the exciation and response is kept at 90 degrees and the excitation is varied, the Resonant decay method (RDM) \cite{Londoo2015}, where the free vibration response of the structure caused by the specific initial conditions is used.

RDM allows independent excitation of modes of the system. The excitation is provided such that the steady-state vibrations are large enough to capture the non-linearity. Once the system starts vibrating at the given excitation, the force is removed and the system exhibits a free decay. The backbone curve is generated using the instantaneous frequency and amplitude of these free vibrations.

\section{Numerical model}\label{Numerical model}
This section explains the extraction of backbone curves of a system for which a mathematical model is provided. In this study, we assume there are no interactions between modes and thus utilize a single degree of freedom system to simplify the analysis and gain a deeper understanding of the system's basic behaviour. This method, predicated on the absence of modal interactions, allows for a fundamental comprehension of the system's dynamics and can illuminate the underlying principles that may apply to more complex systems. Furthermore, employing an SDOF model lowers the computational requirements and enables the utilization of analytical solutions, speeding up the analysis and making it simpler to execute. State space representation of the non-linear system (Equation (\ref{eq:system})) is considered where a set of system parameters are input. 
\begin{equation}
\mathit{m} \ddot{\mathit{x}}+\mathit{c_{1}} \dot{\mathit{x}}+\mathit{k_{1}} \mathit{x}+ \mathit{F}(\mathit{x},\dot{\mathit{x}})=0 \label{eq:system}
\end{equation}

Using RDM, the system is excited at its corresponding first natural frequency which is controlled by giving the initial conditions, (displacement or velocity) in the state space model, large enough to excite the non-linearity. Once excited, the excitation is removed and the free response of the system is collected. \

To determine the initial control conditions, it is crucial to understand the type of non-linearity present in the system. The non-linear function 
${F}({x},\dot{{x}})$ in Equation (2) needs to be characterized, which could be done through empirical experimentation or theoretical analysis. This function governs how the non-linearity behaves with displacement (${x}$) and velocity ($\dot{{x}}$).
The natural frequency in the linear regime is calculated using the mass ($m$) and stiffness ($k$) parameters: $\omega_n= \sqrt{k/m}$. This serves as the initial reference point. Non-linear systems do not have a constant natural frequency like linear systems. The natural frequency of a non-linear system can change with the amplitude of the motion. Therefore, the concept of a 'backbone curve,' which relates the amplitude of oscillation to the instantaneous frequency, is used.\

To determine the initial conditions that excite the first natural frequency, one can start with an estimate based on the linear natural frequency and then adjust the amplitude of the initial condition until the desired non-linear behaviour is observed. If the system is also stochastic, the presence of randomness or noise in the parameters or excitation forces can affect how the initial conditions are set. In this case, a statistical analysis or probabilistic approach may be necessary to determine a range of initial conditions that are most likely to excite the system properly. The fine-tuning of the initial conditions is typically achieved through numerical simulations. The system's response to varying initial conditions is observed, and adjustments are made accordingly. Optimization algorithms may be utilized to streamline this process.

Using this free decay (time response) data, instantaneous frequency and amplitude is extracted using the peak picking method. The sequence of analytical steps required to obtain the backbone curve is described as follows:

\begin{itemize}
    \item Collection of Free Decay Data: Data is collected that captures the free response of the system over time from an initial excited state. 
    \item Identification of Peaks: Within this data, the maximum points of displacement peaks are identified. 
    \item Application of the Peak Picking Method: The period of oscillation is determined by measuring the time intervals between successive peaks, and the amplitude is noted as the magnitude of these peaks. From these periods, the frequencies are calculated, given that frequency is the reciprocal of the period.
    \item Creation of Data Pairs: Each peak corresponds to a data pair consisting of the amplitude of the peak and the frequency at that moment. These pairs form the foundational data points for the backbone curve.
\end{itemize}

\begin{figure}[H]
\begin{center}
\includegraphics[scale= 0.47]{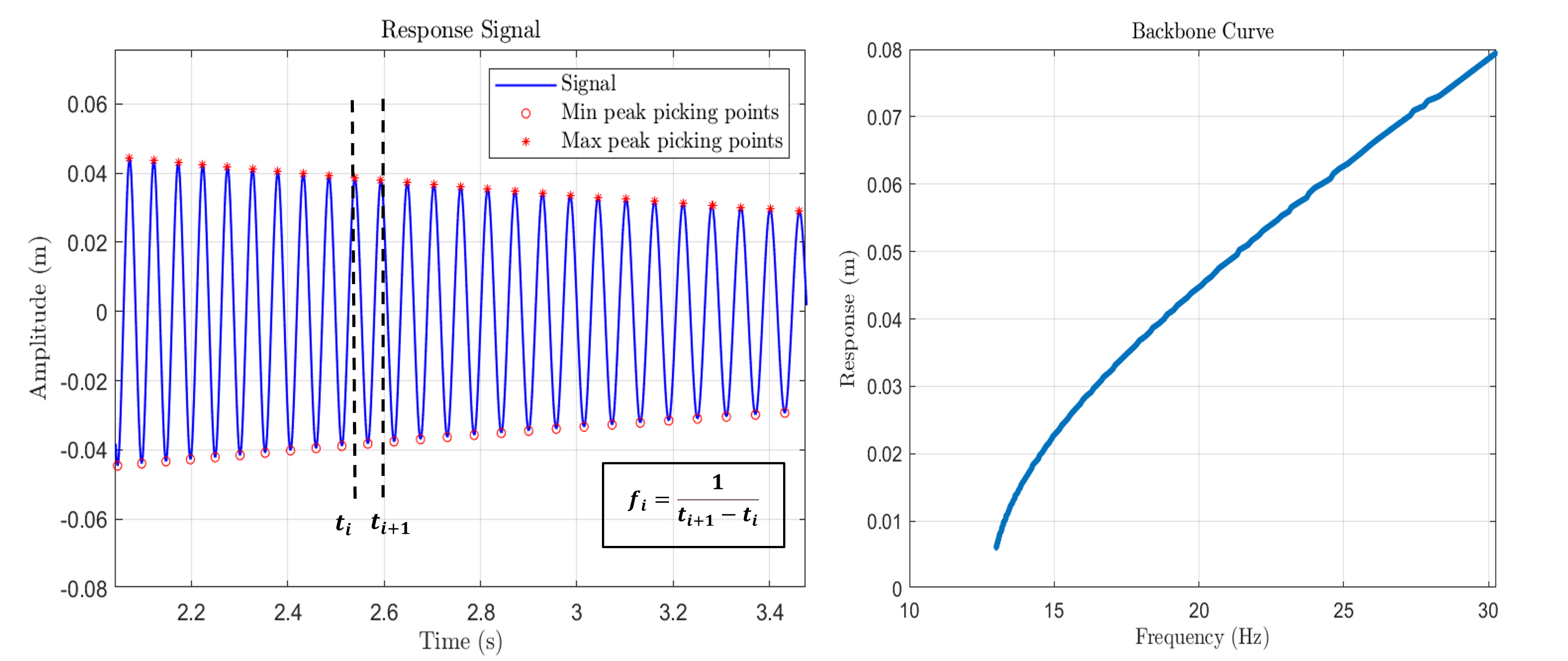}
\caption{Backbone curve}
\label{Backbone}
\end{center}
\end{figure}

These data pairs are plotted on a graph on frequency–amplitude axes. If the system exhibits non-linearity, the frequency may slightly vary with different amplitudes. This resulting curve represents the backbone curve, indicating the system's natural frequency as a function of its oscillation amplitude.
Figure \ref{Backbone} illustrates the process of extracting the backbone curve using the resonant decay method. It shows the relationship between instantaneous frequency and amplitude, providing a graphical representation of the system's dynamic response. This method captures the dynamic behaviour of the system by plotting instantaneous frequency against amplitude, revealing the non-linear characteristics essential for further analysis.

\textcolor{ForestGreen}{The numerical cases (1-5) are formulated to systematically evaluate the effectiveness of the stochastic nonlinear model updating approach across a spectrum of nonlinear behaviours. These cases, ranging from basic nonlinear models to more complex scenarios like dry friction, quadratic damping with cubic stiffness, and the Bouc-Wen model, are designed to test the method's capability in handling various nonlinear phenomena encountered in dynamical systems, including discontinuities, state-dependent forces, and hysteresis.}

Five different non-linear models are considered as test cases for which the mathematical model (Case 1-5) is shown for each case. Each of these cases represents a different possible nonlinear model in dynamical systems. These models, which include dry friction, quadratic damping with cubic stiffness, and the Bouc-Wen model, present unique challenges and insights into the stochastic nonlinear model updating of these systems. The numerical non-linear cases act as foundational tests, establishing the basic efficacy of our approach in a well-controlled setting with clear non-linear characteristics. The dry friction case delves into the non-linearities induced by frictional forces, common in mechanical systems, challenging our method to handle discontinuities and variable force behaviours typical in engineering. The quadratic damping with cubic stiffness scenario adds complexity by introducing damping forces that vary with velocity squared and stiffness that changes with displacement cubed, testing our approach's capability with state-dependent non-linear restoring and damping forces.\

The Bouc-Wen model incorporates an advanced hysteresis non-linearity, critical for systems experiencing cyclic loading and energy dissipation, such as in seismic structural responses. To gain a comprehensive understanding of the Bouc-Wen model, its intricacies, and practical implementations, interested readers are encouraged to consult the detailed analyses and case studies presented in references  \cite{Baber1985, Charalampakis2008, Ikhouane2007}. In Case 5, we address a scenario in which the mathematical model of the system is incorrectly chosen, a situation referred to as incorrect modelling assumptions in traditional model updating. In this scenario, the measured data are simulated using the Bouc-Wen model, while the model assumed for updating is a system with nonlinear springs and dampers. This approach reflects real-world scenarios where modelling errors might occur. Here, the measured data are employed to identify the parameters of the assumed model, thereby improving the correlation between the model predictions and the measured data.\

\begin{figure}[H]
\subfigure[Cubic stiffness with $k_{1}$ = 6500 N/m, $k_{2}$ = 6.5 \textnormal{MN/m$^{3}$}, $c_{1}$ = 0.8 Ns/m]{\includegraphics[width= \textwidth]{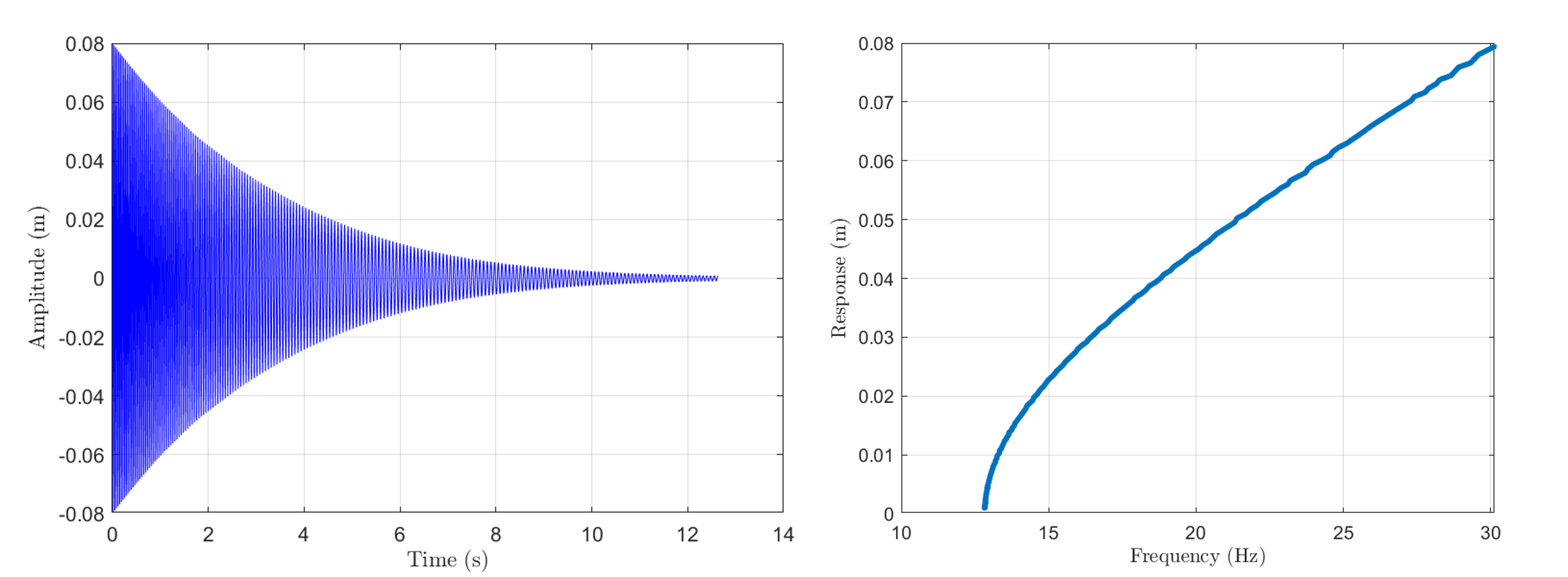}}
\subfigure[Dry friction with $k_{1}$ = 6500 N/m, $c_{1}$ = 0.8 Ns/m, $c_{2}$ = 0.3 N]{\includegraphics[width= \textwidth]{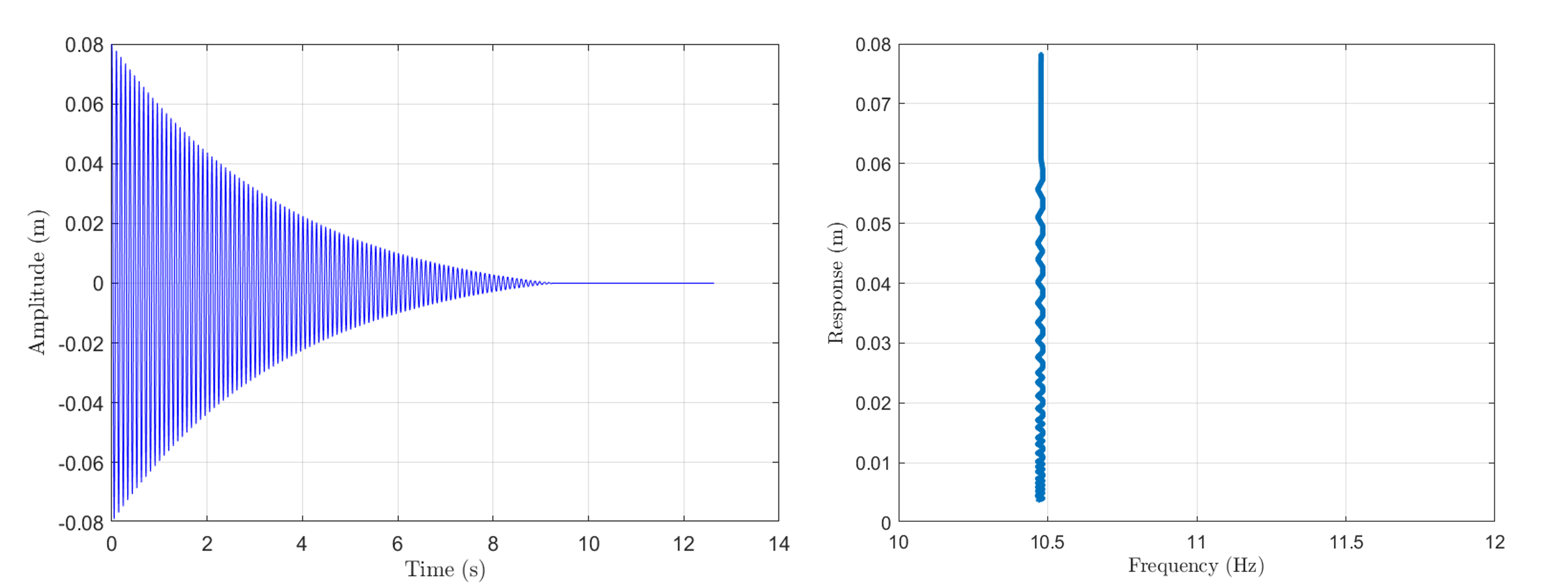}}
\subfigure[Quadratic damping with cubic stiffness with $k_{1}$ = 6500 N/m, $k_{2}$ = 6.5\textnormal{MN/m$^{3}$},  $c_{1}$ = 0.8 Ns/m, $c_{2}$ = 0.3 \textnormal{Ns$^2$/m$^2$}]{\includegraphics[width= \textwidth]{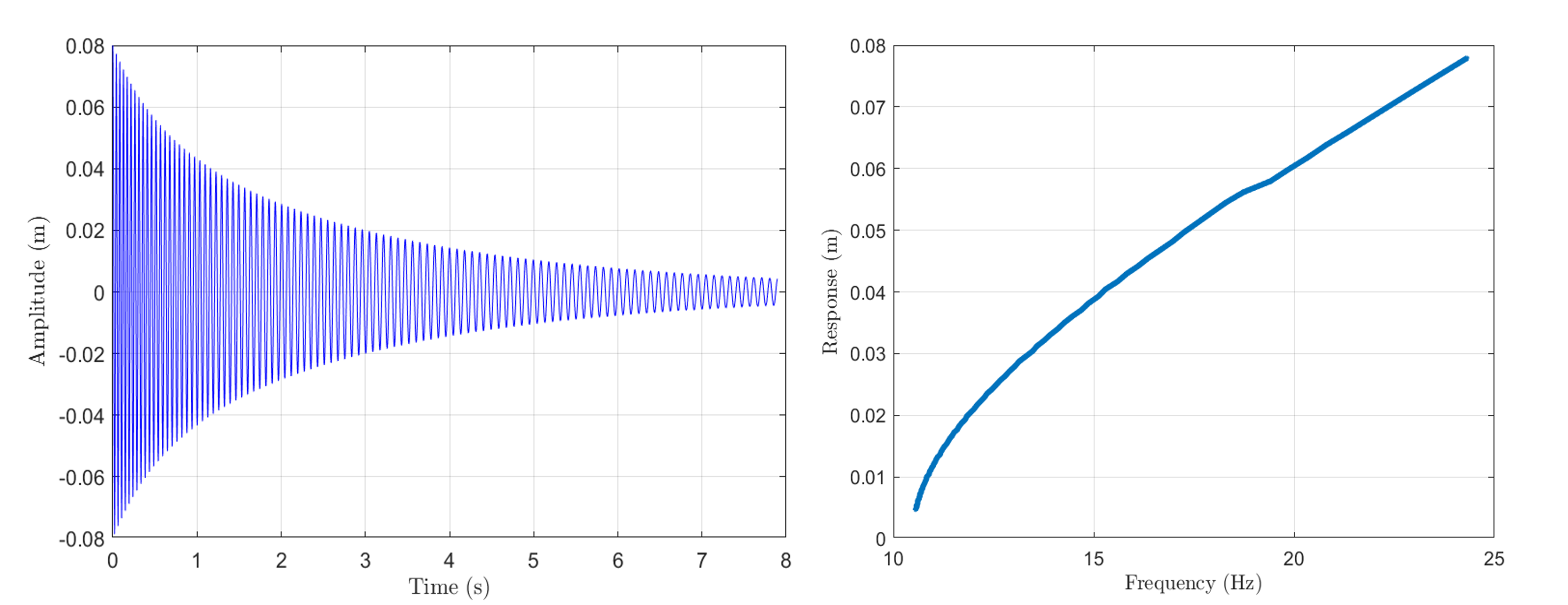}}
\subfigure[Bouc-Wen model with $A$ = 1.95, $\alpha$ = 0.79 , $\beta$ = 1.97  m$^{-1}$, $\gamma$ = 1.98 m$^{-1}$]{\includegraphics[width= \textwidth]{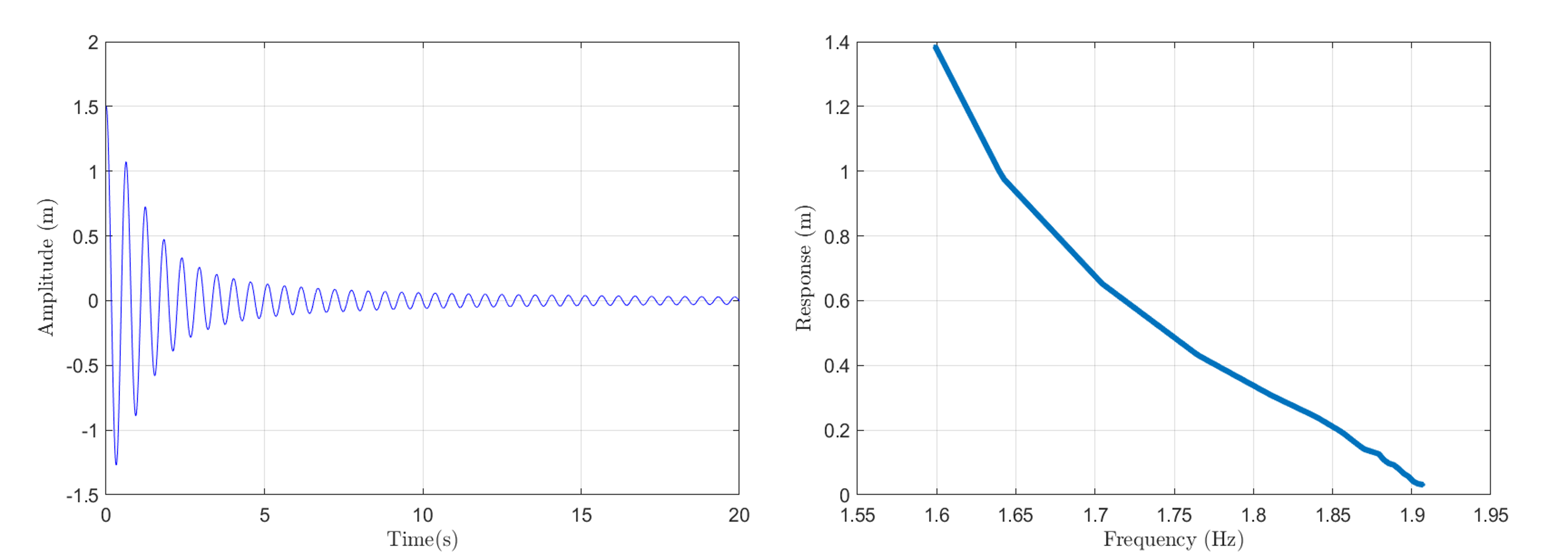}}
\caption{Backbone curves of the cases using the response decay method}
\label{backbones}
\end{figure}

Figure \ref{backbones} presents estimated backbone curves for the five distinct non-linear models: cubic stiffness, dry friction, quadratic damping with cubic stiffness and the Bouc-Wen model. Case 5 employs the Bouc-Wen model's backbone curves for stochastic parameter identification. Each plot in the figure corresponds to one of these numerical cases, visually showing how the system's dynamic behaviour changes under different non-linear conditions. Together, these plots illustrate the versatility of the backbone curve in capturing a wide range of non-linear dynamics. 
The equations of motion representing the non-linearity in all five test cases are:\\\\
\textbf{Case 1:} Cubic stiffness, $m\ddot{x} + c_1\dot{x} + k_1 x + k_2 x^3 = 0$ \\
\textbf{Case 2:} Dry friction, $m\ddot{x} + c_1\dot{x} + k_1 x + c_2 \operatorname{sign}(\dot{x}) = 0$ \\
\textbf{Case 3:} Quadratic damping with cubic stiffness, $m\ddot{x} + c_1\dot{x} + k_1 x + c_2 \dot{x}|\dot{x}| + k_2 x^3 = 0$ \\
\textbf{Case 4:} Bouc-Wen model, 
\( m\ddot{x} + c_1\dot{x} + k_1 x + (1 - \alpha)kz = 0, \) \\
where \( \dot{z} = A \dot{x} - \beta | \dot{x} || z^{n-1} |z- \gamma \dot{x} |z|^n \)\\
\textbf{Case 5:} Approximated data-driven model/ Empirical model, $m\ddot{x} + c_1\dot{x} + k_1 x + k_2 x^2 + c_2 |\dot{x}|\dot{x} + k_3 x^3 + c_3 \dot{x}^3$\\

The estimated backbone curves represent the system's possible measurements. In this case, the inverse problem, in which the parameters of interest are calculated based on the observations, is solved using Bayesian inference.

\subsection{Results from MCMC sampling for numerical cases}

\begin{figure}[H]
\begin{center}
\includegraphics[scale= 0.45]{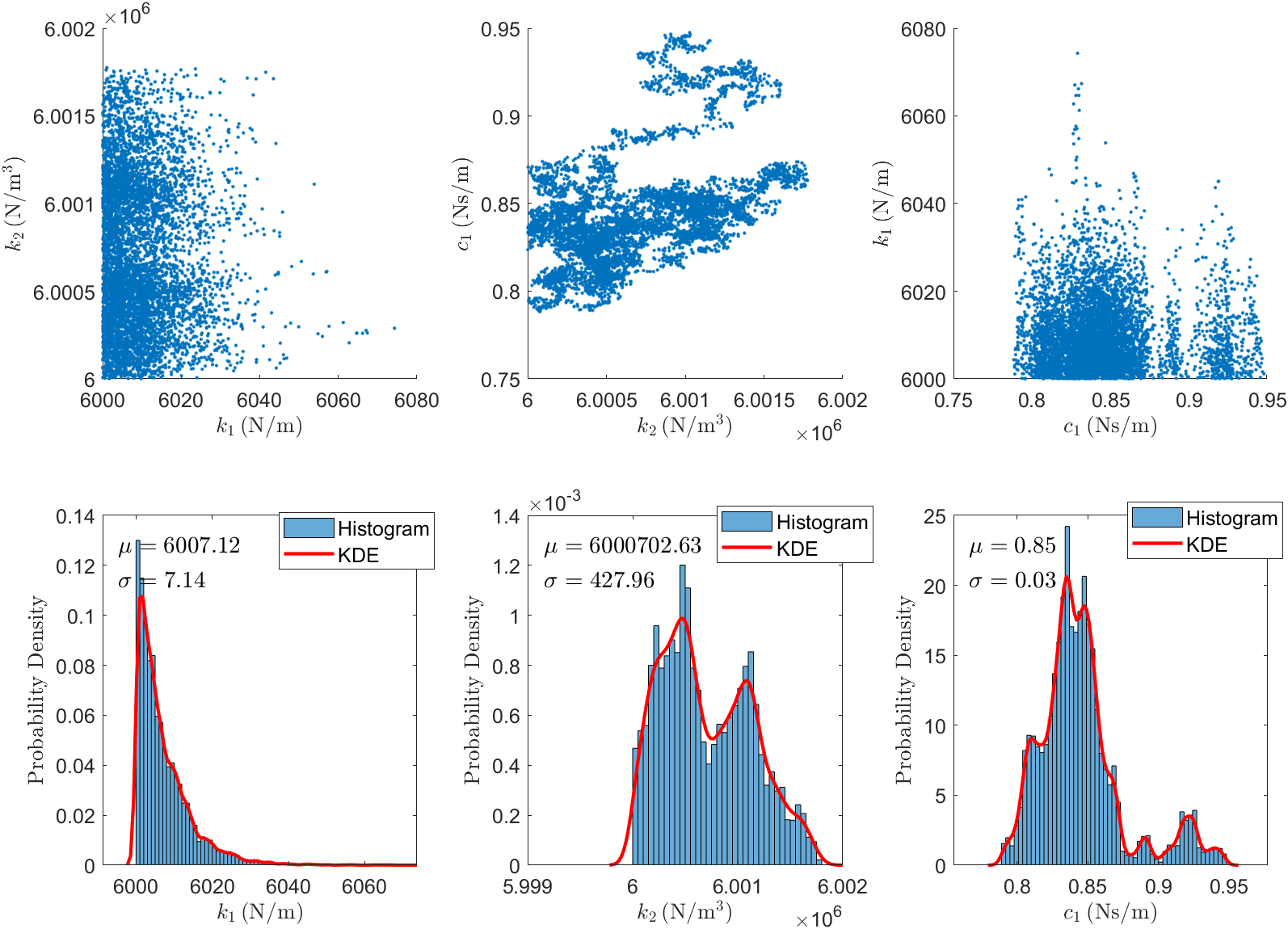}
\caption{MCMC Sampling Results: Histogram and Kernel Density Estimation (KDE) for parameters $k_{1}$, $k_{2}$ and $c_{1}$ for Case 1.}
\label{fig:singleValue}
\end{center}
\end{figure}

\textcolor{ForestGreen}{To validate the methodology, a fundamental verification was first performed using single target values considering the cubic stiffness case for the parameters ($k_1 = 6000 \, \text{N/m}$, $k_2 = 6.0 \times 10^6 \, \text{N/m}^3$, $c_1 = 0.8 \, \text{Ns/m}$). The MCMC sampling results, shown in Figure \ref{fig:singleValue}, demonstrate that the method successfully converges to these target values, producing non-uniform posterior distributions centred around the true parameters.}


\textcolor{ForestGreen}{This single-value validation confirms the basic effectiveness of the updating framework in identifying specific parameter values. Building upon this foundation, the approach was extended to handle parameter ranges, presenting a more challenging test case as the method requires capturing not just specific values but entire probability distributions.}

In stochastic modelling via MCMC sampling for the five test cases, the selection of parameters for each model is based on capturing the distinctive non-linear characteristics inherent in each system. For Cases 1-3, the focus is on linear and non-linear stiffness ($k_{1}$, $k_{2}$) and damping ($c_{1}$, $c_{2}$) parameters. These parameters are crucial for modelling the non-linear response due to stiffness and damping effects, which are predominant features in these scenarios. In the Bouc-wen model, parameters such as \( A\), \(\alpha \), \( \beta\) and \(\gamma\) are critical for representing the hysteresis non-linearity. The Bouc-Wen model's complexity and its ability to simulate hysteresis in structural systems necessitate these specific parameters to accurately capture the non-linear hysteretic behaviour.\

\begin{figure}[H]
\begin{center}
\includegraphics[scale= 0.75]{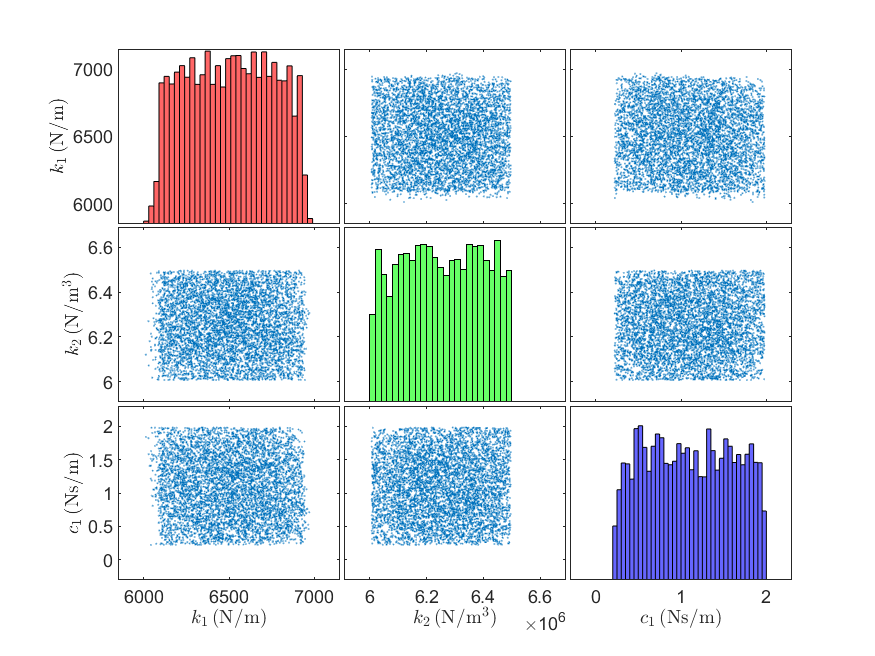}
\caption{Distribution of 20000 MCMC samples of $k_{1}$ (Red), $k_{2}$ (Green)and $c_{1}$ (Blue) for Case 1.}
\label{20000scatterplot_duff}
\end{center}
\end{figure}

\begin{figure}[H]
\begin{center}
\includegraphics[scale= 0.75]{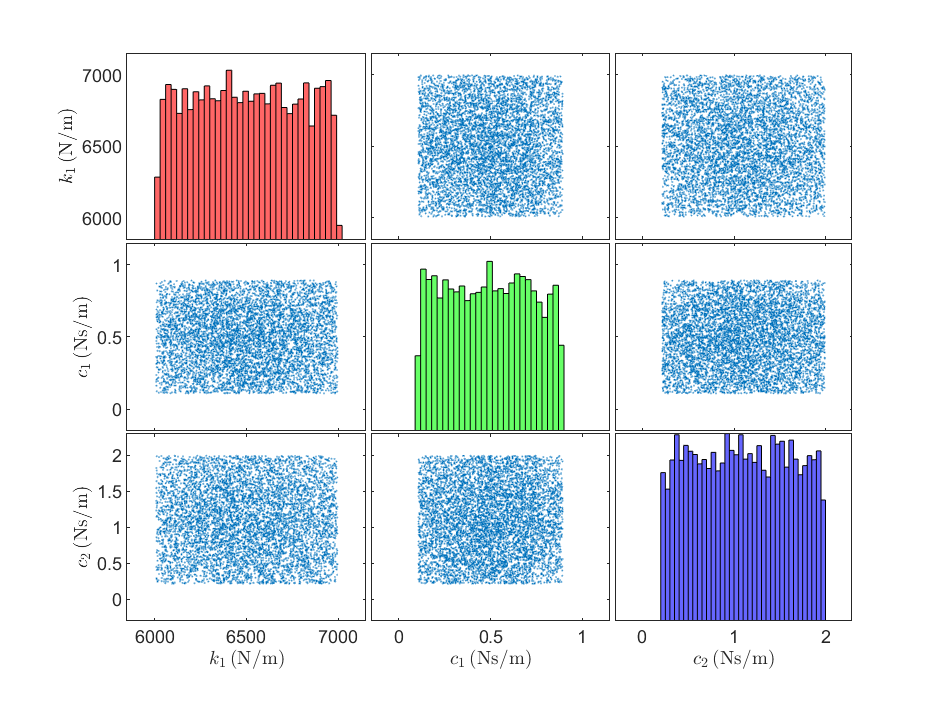}
\caption{Distribution of 20000 MCMC samples of $k_{1}$ (Red), $c_{1}$ (Green)and $c_{2}$ (Blue) for Case 2.}
\label{20000scatterplot_dry}
\end{center}
\end{figure}

\begin{figure}[H]
\centering
\includegraphics[scale= 0.44]{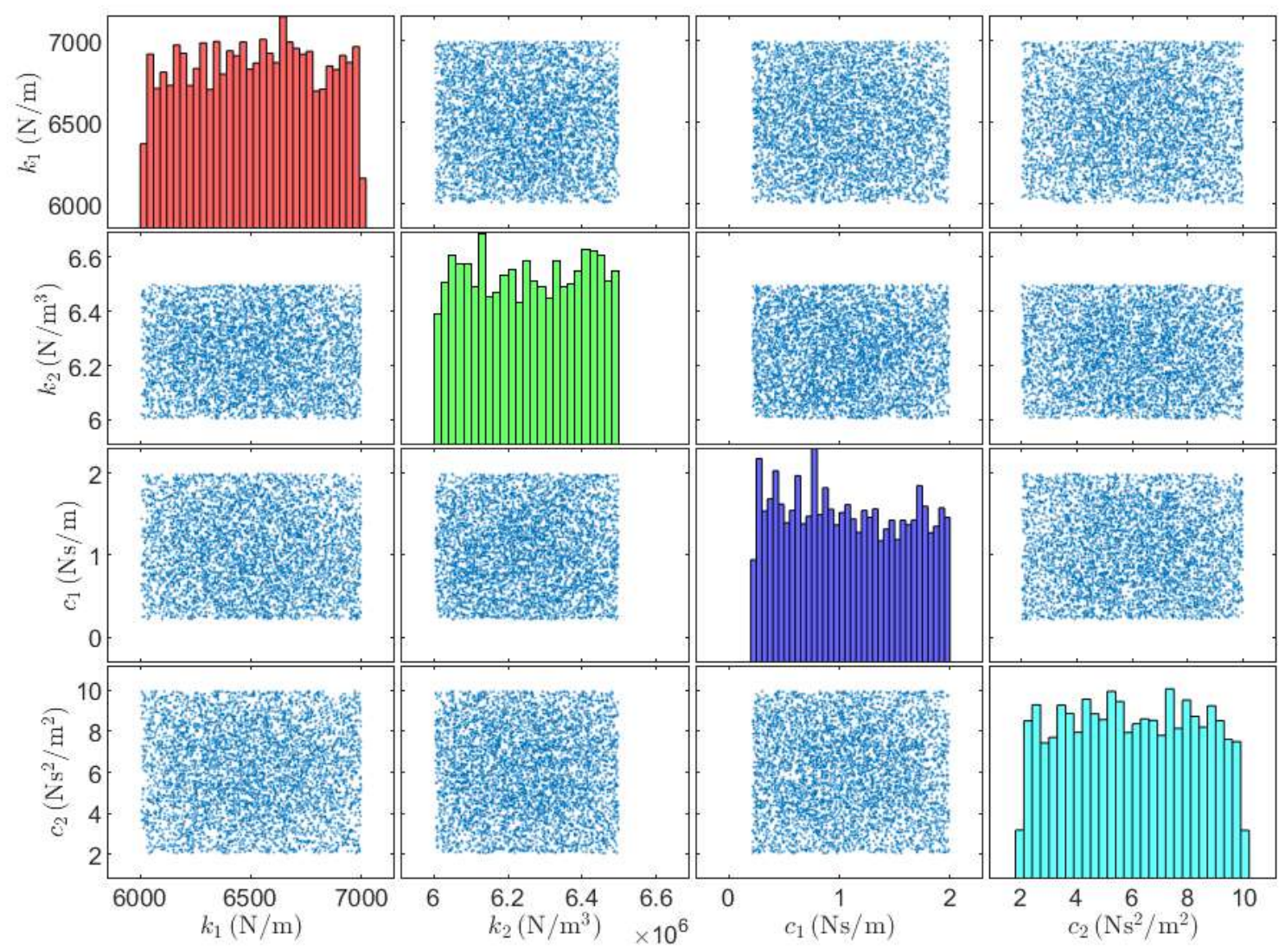}
\caption{Distribution of 20000 MCMC samples of $k_{1}$ (Red), $k_{2}$ (Green), $c_{1}$ (Blue) and $c_{2}$ (Steel blue) for Case 3.}
\label{20000scatterplot_quad}
\end{figure}

\begin{figure}[H]
\begin{center}
\includegraphics[scale= 0.45]{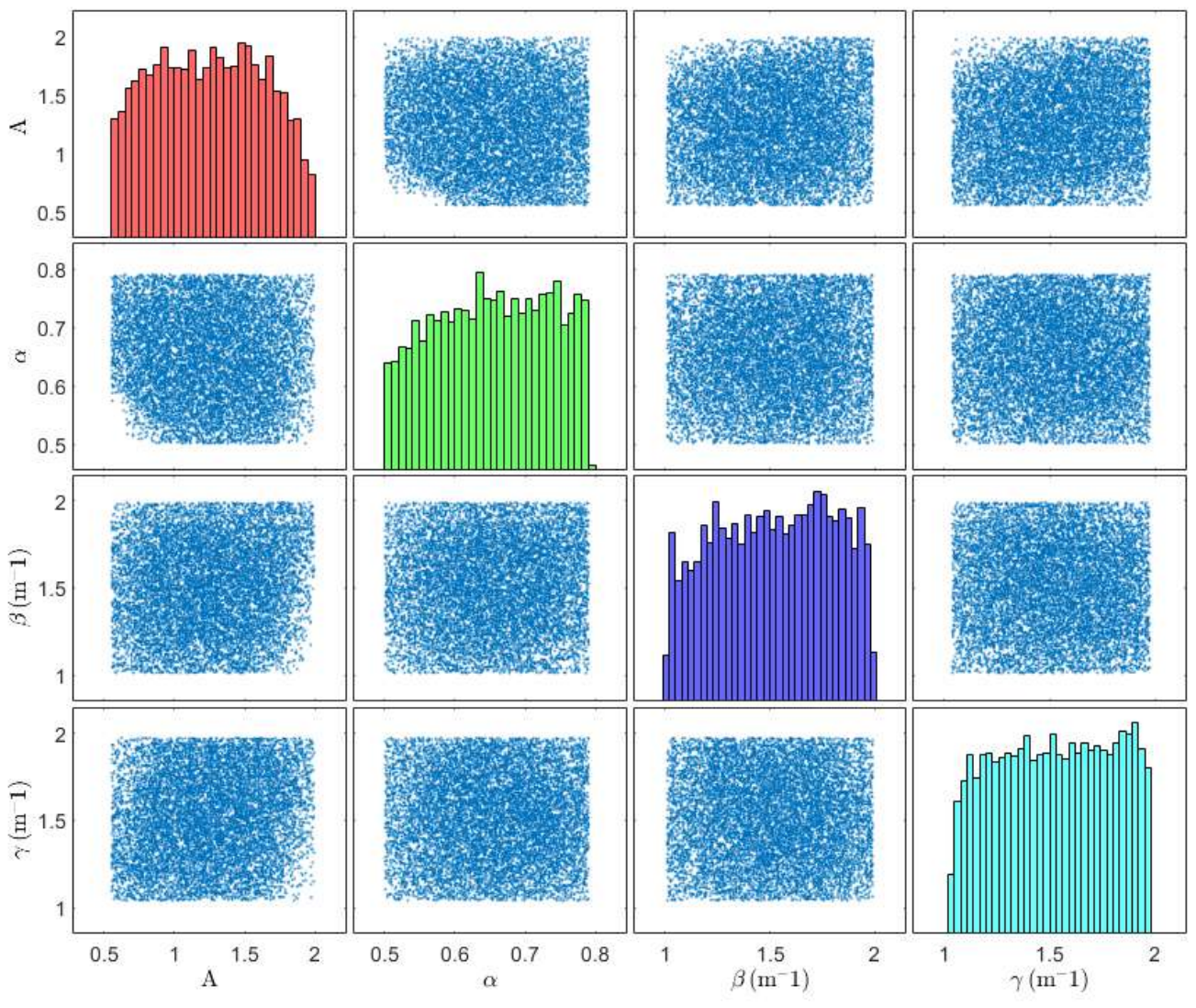}
\caption{Distribution of 20000 MCMC samples of \( A \) (Red), \( \alpha \) (Green), \( \beta \) (Blue) and \(\gamma \) (cyan) for Case 4.}
\label{20000scatterplot_bouc}
\end{center}
\end{figure}

\begin{figure}[H]
\begin{center}
\includegraphics[width=1.1\textwidth]
{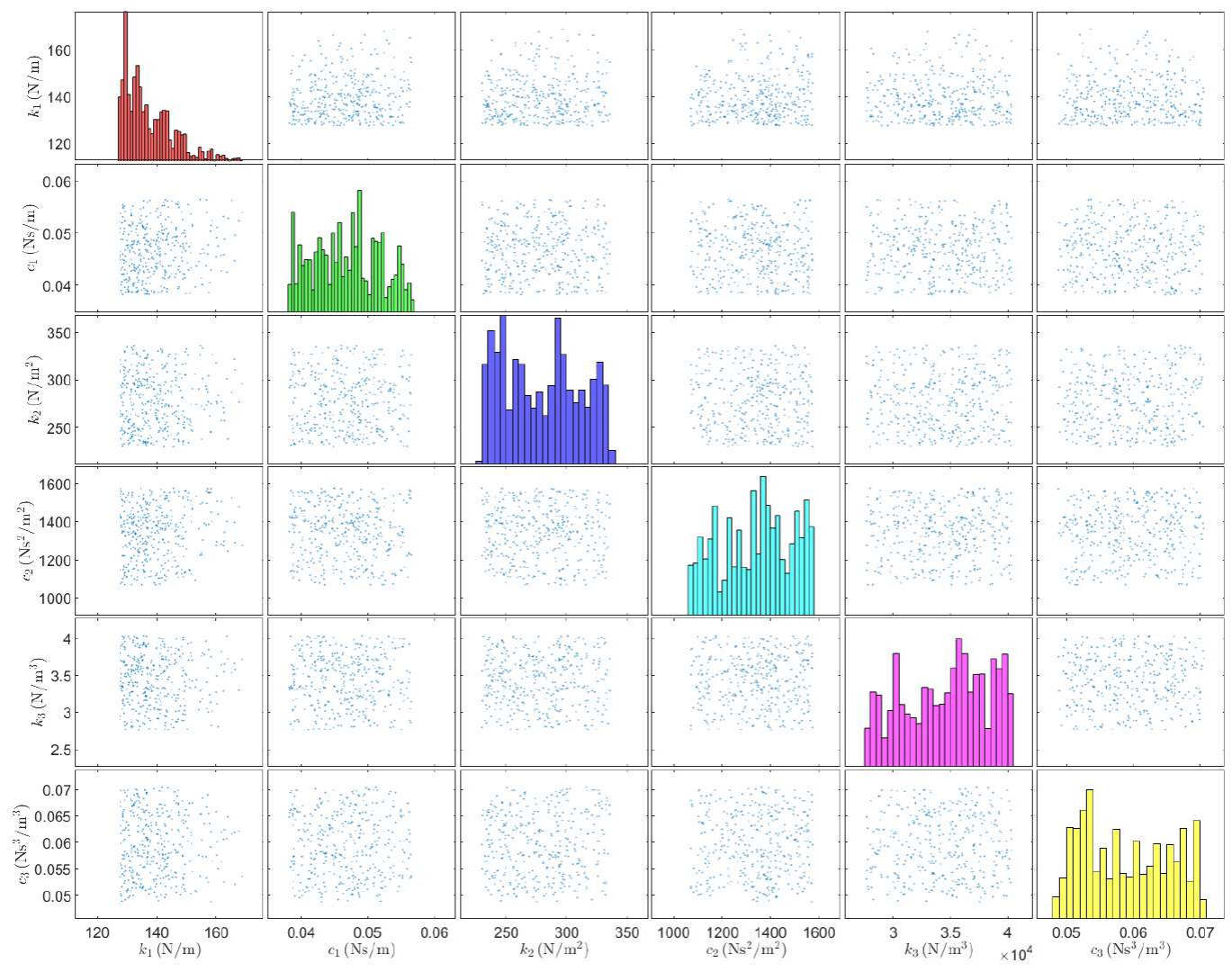}
\caption{Distribution of 20,000 MCMC samples of $k_{1}$ (Red), $c_{1}$ (Green),  $k_{2}$ (Red), $c_{2}$ (Cyan), $k_{3}$ (Magenta), and $c_{3}$ (Yellow) for Case 5.}
\label{20000scatterplot_approx}
\end{center}
\end{figure}

 The range of parameters for both the true and prior distributions in all the cases is specified in Tables \ref{table:trueprior_duff}, \ref{table:trueprior_dry}, \ref{table:trueprior_quad}, \ref{table:trueprior_bouc} and \ref{table:trueprior_approx}. The distribution of samples corresponding to model parameters can be observed in the accompanying figures \ref{20000scatterplot_duff}, \ref{20000scatterplot_dry}, \ref{20000scatterplot_quad}, \ref{20000scatterplot_bouc}, \ref{20000scatterplot_approx}.\\

\textbf{Case 1:} Cubic stiffness
\begin{table}[ht]
\centering
\small
\setlength{\tabcolsep}{10pt}
\caption{Parameter Statistics and MCMC Results for Case 1}
\label{table:trueprior_duff}
\begin{tabular}{@{}p{1.8cm}p{2cm}p{2cm}p{2.7cm}@{}}
\hline
Parameter & True \newline Mean ($\mu$), Std Dev ($\sigma$) & Prior \newline Mean ($\mu$), Std Dev ($\sigma$) & MCMC samples \newline Mean ($\mu$), Std Dev ($\sigma$) \\
\hline
$k_{1}$ (N/m) & 6500, 288.68 & 6550, 664.56 & 6507.74, 245.15 \\
$k_{2}$ (MN/m$^{3}$) & 6.25, 0.14 & 6.2, 0.46 & 6.25, 0.14 \\
$c_{1}$ (Ns/m) & 1.1, 0.52 & 1.55, 0.85 & 1.10, 0.5 \\
\hline
\end{tabular}
\end{table}


Parameters considered are $k_{1}$, $k_{2}$ and $c_{1}$ for the cubic stiffness case. Table \ref{table:trueprior_duff} shows the wider range of prior given for MCMC sampling and the true (actual) range of parameters. The range and distribution for  $k_{1}$, $k_{2}$, and $c_{1}$ (Figure \ref{20000scatterplot_duff}) are near to its true distribution with an acceptance rate of 0.33.  Figure \ref{20000scatterplot_duff} showcases the distribution of 20,000 MCMC samples for the parameters  $k_{1}$ and $k_{2}$ in Case 1 of the study. This visualization aids in understanding the spread and correlation of these parameters, highlighting the effectiveness of the MCMC sampling approach in parameter estimation. However, the distribution is uniform with a range similar to the one defined for true distribution.\\

\textbf{Case 2:} Dry friction
\begin{table}[ht]
\centering
\small
\setlength{\tabcolsep}{10pt}
\caption{Parameter Statistics and MCMC Results for Case 2}
\label{table:trueprior_dry}
\begin{tabular}{@{}p{1.8cm}p{2cm}p{2cm}p{2.7cm}@{}}
\hline
Parameter & True \newline Mean ($\mu$), Std Dev ($\sigma$) & Prior \newline Mean ($\mu$), Std Dev ($\sigma$) & MCMC samples \newline Mean ($\mu$), Std Dev ($\sigma$) \\
\hline 
$k_{1}$ (N/m) & 6500, 288.68 & 6550, 664.56 & 6507.75, 280.13 \\
$c_{1}$ (Ns/m) & 0.5, 0.23 & 2.55, 1.41 & 0.49, 0.23  \\
$c_{2}$  (N) & 1.1, 0.52 & 5.05, 2.85 & 1.09, 0.51  \\
\hline
\end{tabular}
\end{table}


Figure \ref{20000scatterplot_dry} illustrates the distribution of 20,000 MCMC samples for parameters $k_{1}$, $c_{1}$ and $c_{2}$ in Case 2. The distribution patterns provide insights into the behaviour of these parameters under the given conditions and the reliability of the MCMC sampling method used for this case. The distribution of parameters is uniform and is within the range of true distribution as can be seen from Table \ref{table:trueprior_dry}. The MCMC samples have an acceptance rate of 0.34.\\

\textbf{Case 3:} Quadratic damping with cubic stiffness
\begin{table}[ht]
\centering
\small
\setlength{\tabcolsep}{10pt}
\caption{Parameter Statistics and MCMC Results for Case 3}
\label{table:trueprior_quad}
\begin{tabular}{@{}p{1.8cm}p{2cm}p{2cm}p{2.7cm}@{}}
\hline
Parameter & True \newline Mean ($\mu$), Std Dev ($\sigma$) & Prior \newline Mean ($\mu$), Std Dev ($\sigma$) & MCMC samples \newline Mean ($\mu$), Std Dev ($\sigma$) \\
\hline
$k_{1}$ (N/m) & 6500, 288.68 & 6550, 664.56 & 6511.17, 281.26 \\
$k_{2}$ (\textnormal{MN/m$^3$}) & 6.25, 0.14 & 6.275, 0.50 & 6.25, 0.144 \\
$c_{1}$ (Ns/m) & 1.1, 0.52 & 5.05, 2.85 & 1.08, 0.52 \\
$c_{2}$ (\textnormal{Ns$^2$/m$^2$}) & 6, 2.31 & 10.5, 5.48 & 5.98, 2.27 \\
\hline
\end{tabular}
\end{table}


In Figure \ref{20000scatterplot_quad}, the distribution of 20,000 MCMC samples for parameters $k_{1}$, $k_{2}$, $c_{1}$ and $c_{2}$ in Case 3 is displayed. This figure helps in analyzing the variability and relationship among these parameters, showcasing the applicability and efficiency of the MCMC method in exploring the parameter space for this specific case.
The distribution of the four parameters of the system  ($k_{1}$, $k_{2}$, $c_{1}$ and $c_{2}$) in Case 3 appear to be uniformly distributed, which aligns with the actual distribution shown in Table \ref{table:trueprior_quad}. The acceptance rate of MCMC samples generated is 0.23.\\

\textbf{Case 4:} Bouc-Wen model
\begin{table}[ht]
\centering
\small
\setlength{\tabcolsep}{10pt}
\caption{Parameter Statistics and MCMC Results for Case 4 (Bouc-Wen Model)}
\label{table:trueprior_bouc}
\begin{tabular}{@{}p{1.8cm}p{2cm}p{2cm}p{2.7cm}@{}}
\hline
Parameter & True \newline Mean ($\mu$), Std Dev ($\sigma$) & Prior \newline Mean ($\mu$), Std Dev ($\sigma$) & MCMC samples \newline Mean ($\mu$), Std Dev ($\sigma$) \\
\hline
\( A \)  & 1.25, 0.43 & 2.05, 1.13 & 1.25, 0.38 \\
\( \alpha \)  & 0.65, 0.09 & 0.45, 0.20 & 0.65, 0.08 \\
\( \beta \) (m$^{-1}$) & 1.5, 0.29 & 1.25, 0.43 & 1.52, 0.28 \\
\(\gamma \) (m$^{-1}$)  & 1.5, 0.29 & 1.75, 0.72 & 1.52, 0.27 \\
\hline
\end{tabular}
\end{table}


Figure \ref{20000scatterplot_bouc} presents the uniform distribution of 20,000 MCMC samples of parameters of the Bouc-Wen model (\( A\), \(\alpha \), \( \beta\) and \(\gamma\)), which is in agreement with the actual distribution given in Table \ref{table:trueprior_bouc}. The variance in the samples reflects the uncertainty in parameter estimates. The acceptance rate for these generated MCMC samples is 0.5.\\

\textbf{Case 5:} Approximated data-driven model/ Empirical model\\

In Case 5, backbone curves from the Bouc-Wen model are utilized to infer parameters of an empirical model through MCMC sampling, where direct equations are not available. The true parameter ranges are deduced from the Bouc-Wen model's time series data, and the MCMC samples determine the identified range. This approach underscores the adaptability of MCMC methods in system identification, even when explicit model equations are absent, by leveraging measured data to capture system dynamics.
\begin{table}[ht]
\centering
\small
\setlength{\tabcolsep}{10pt}
\caption{Parameter Statistics and MCMC Results for Case 5}
\label{table:trueprior_approx}
\begin{tabular}{@{}p{2cm}p{2cm}p{2cm}p{2.7cm}@{}}
\hline
Parameter & True \newline Mean ($\mu$), Std Dev ($\sigma$) & Prior \newline Mean ($\mu$), Std Dev ($\sigma$) & MCMC samples \newline Mean, Std Dev ($\sigma$) \\
\hline 
$k_{1}$ (N/m) & 156.5, 18.17 & 150, 11.55 & 137.28, 8.26 \\
$c_{1}$ (Ns/m) & 0.0465, 0.0055 & 0.048, 0.0046 & 0.05, 0.01 \\
$k_{2}$ (\textnormal{N/m$^2$}) & 283.5, 32.74 & 285, 37.53 & 279.33, 31.57 \\
$c_{2}$ (\textnormal{kNs$^2$/m$^2$}) & 1.25, 0.14 & 1.3, 0.17 & 1.34, 0.15 \\
$k_{3}$ (\textnormal{kN/m$^3$}) & 33.5, 3.75 & 32.5, 4.33 & 34.66, 3.65 \\
$c_{3}$ (\textnormal{Ns$^3$/m$^3$}) & 0.059, 0.0069 & 0.06, 0.0058 & 0.06, 0.01 \\
\hline
\end{tabular}
\end{table}


In Figure \ref{20000scatterplot_approx}, the distribution of 20,000 MCMC samples for parameters $k_{1}$,$c_{1}$ $k_{2}$, $c_{2}$, $k_{3}$, and $c_{3}$ in Case 5 is displayed. This figure helps in analyzing the variability and relationship among these parameters, showcasing the applicability and efficiency of the MCMC method in exploring the parameter space for this specific case.
 An acceptance rate of 0.02 was observed for MCMC sampling of six parameters suggesting a very low proportion of proposed samples being accepted, indicating a further need to tune the proposal distribution to the target distribution. However, the parameters ($k_{1}$,$c_{1}$ $k_{2}$, $c_{2}$, $k_{3}$, and $c_{3}$) still manage to reflect the true range (shown in Table \ref{table:trueprior_approx}) within the system's variability.\
 
 When attempting to fit a model to data that adheres to the Bouc-Wen model's complexity, utilizing a less complex mathematical representation can result in significant estimation errors. The Bouc-Wen model's intricacies, specifically its non-linear hysteresis behaviour, are integral to its dynamics, which simpler models may fail to encapsulate. This mismatch often leads to a lower acceptance rate in MCMC sampling due to inadequate representation of the system's true behaviour. Recognizing and addressing these challenges—whether they stem from model selection errors or parameter estimation inaccuracies—is critical for the fidelity of the model to the physical system it seeks to represent.\
 
 \textcolor{ForestGreen}{From Tables 1-5, it can be inferred that the MCMC results successfully recover the true values of all system parameters with reduced uncertainty (lower standard deviations) compared to the prior estimates. MCMC process effectively refined the initial prior estimates, particularly in terms of reducing uncertainty and aligning the mean values with the true values.} The scatterplot of cases 1-4 suggests that the samples are uniformly distributed with a range that resembles the one defined for the true distribution. However, for Case 5, although the range of parameters has been identified, more MCMC samples would be required to capture the true parameter distribution. \textcolor{ForestGreen}{From these tables and their corresponding scatterplots for each case, it is evident that the MCMC samples capture not only the true parameter statistics but also reflect the true distribution, which was uniform for all numerical cases.}

 \textcolor{ForestGreen}{Cases 1-4 are known numerical cases and for these known numerical cases, we intentionally generated backbone curves using uniformly sampled parameters. This approach was chosen to verify the MCMC method's ability to accurately sample from a known target distribution. The uniformity in the posteriors demonstrates that our method successfully reconstructs the original parameter distribution, validating the core functionality of our MCMC implementation. In contrast to Cases 1-4, Case 5 exhibits non-uniform posterior distributions. This case represents a more realistic scenario where we approximated an unknown complex model (Bouc-Wen) with a simpler empirical model. The resulting non-uniform distributions reflect the uncertainty introduced by this model approximation, showcasing our method's behaviour when dealing with model form uncertainty.
 While Cases 1-4 don't demonstrate parameter estimation in the traditional sense, they serve as a crucial verification step, confirming that our MCMC implementation correctly samples from the target distribution.  Case 5 scenario provides a more practical test of our method's capabilities in handling real-world uncertainties and model approximations.} 
 
 The trace plots of parameters show the performance of the algorithm with the given proposal distribution. The trace plot for each parameter for each case can be seen in the Appendix.

\textcolor{ForestGreen}{The computational efficiency of the MCMC sampling process was notably high, primarily due to the straightforward nature of the nonlinear systems under investigation. For the initial three test cases (cases 1-3), the simulation successfully generated 20,000 samples within approximately 15 minutes. The more complex cases 4 and 5 required slightly more processing time, taking roughly 21 minutes to complete the same number of samples. This sampling was optimized through parallel processing techniques, specifically utilizing a quad-core configuration to distribute the computational load. The experimental case demonstrated particular efficiency since it involved a cubic stiffness system - a relatively simple nonlinear form - which contributed to its faster processing time. The implementation of parallel computing architecture played a crucial role in achieving these efficient processing times, as it allowed for simultaneous execution of multiple sampling chains across the four available processing cores.}
 
 \subsection{Comparison of MCMC results of Bouc-Wen and approximated model}
\begin{figure}[H]
\centering
\includegraphics[width = 60mm]{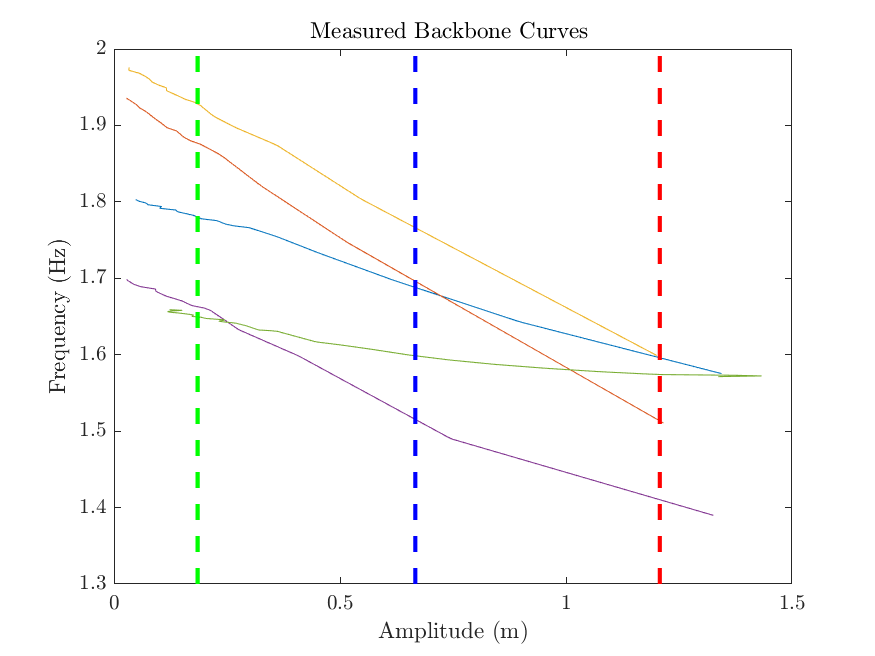}
\includegraphics[width = 60mm]{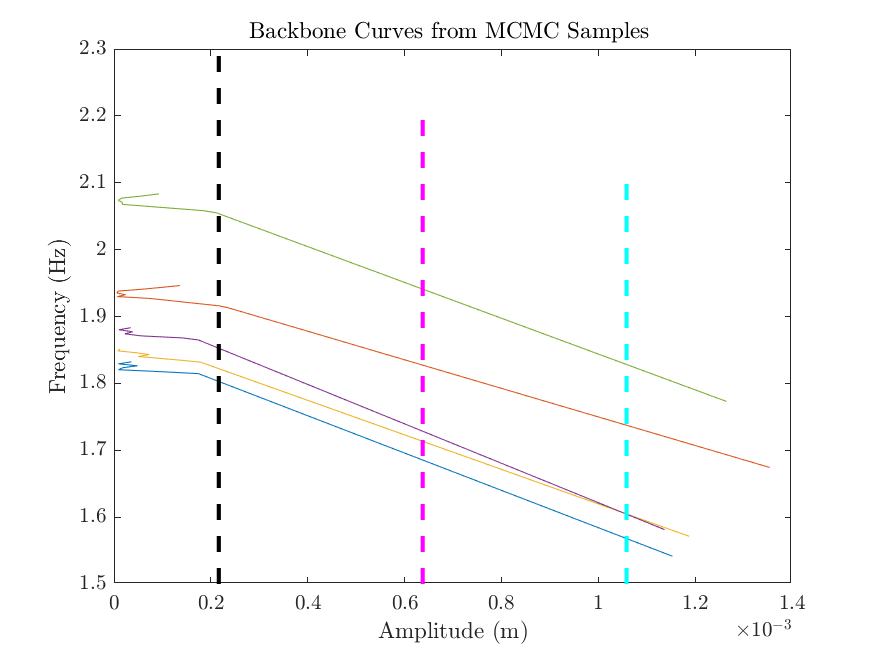}
\caption{Measured (Bouc-Wen model) and MCMC sampled backbone curves}
\label{pdf_bb_curve}
\end{figure}

 A comparison was made between probability density functions (PDFs) generated from two different sources: one set from the experimental model, represented by the Bouc-Wen model's Backbone curves, and another set derived from the identified approximated stochastic nonlinear model in case 5. The main purpose of this case study is to assess the effects of modelling errors on the identification of the statistical properties of the output. This comparison is depicted for three amplitude levels: low, medium, and high.
 Figure \ref{pdf_bb_curve} showcases two sets of Backbone Curves, the left plot displays the measured Backbone Curves, which are derived from the Bouc-Wen model, while the right plot displays the Backbone Curves obtained from MCMC sampled data, which represent the approximated model's output.

\begin{figure}[H]
\centering
\includegraphics[width = \linewidth]
{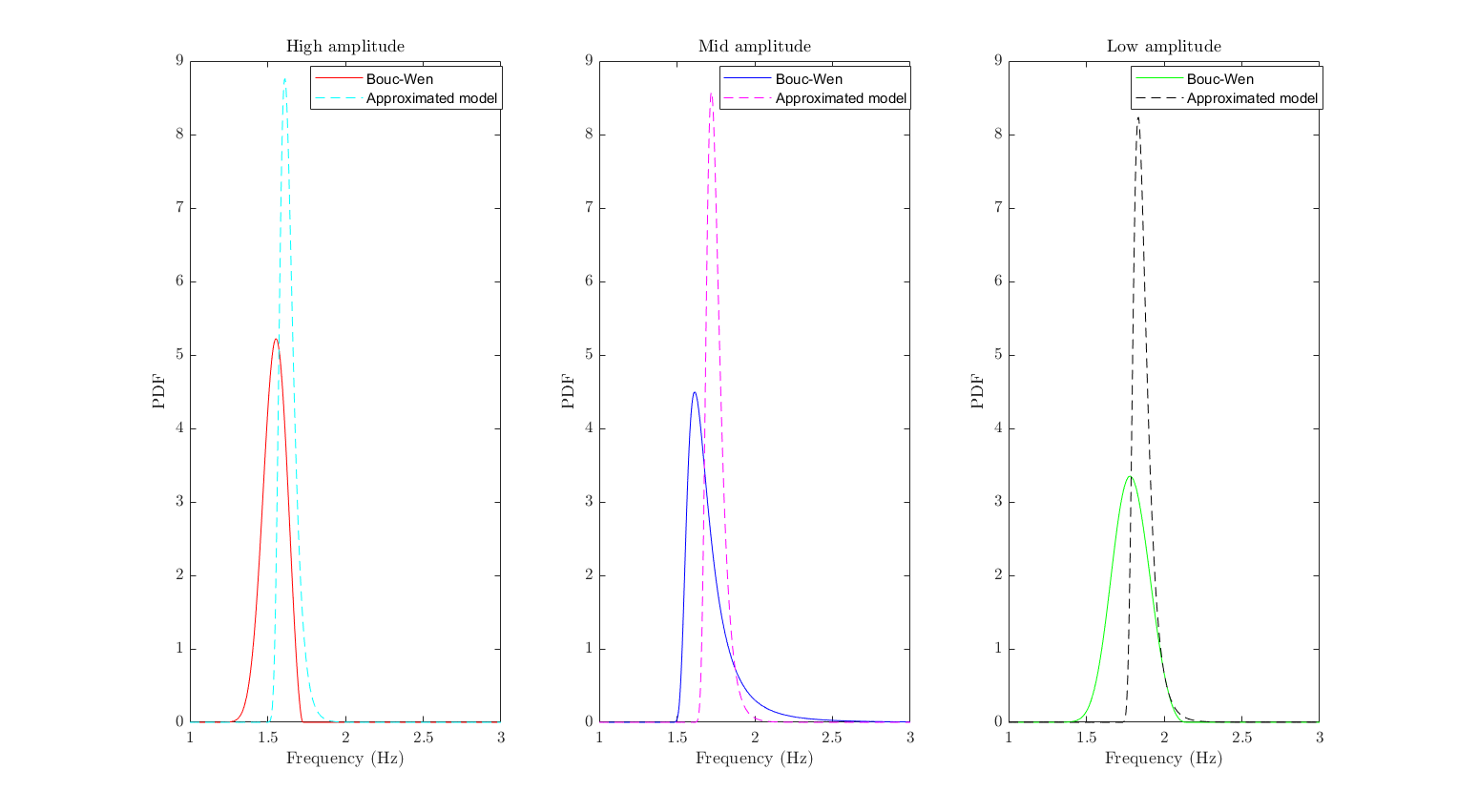}
\caption{PDF Comparison of MCMC results of Bouc-Wen and approximated model}
\label{pdf_comp_Bouc}
\end{figure}
The dashed lines in contrasting colours mark specific amplitude levels (denoted in red, green, and blue for the Bouc-Wen model, and cyan, magenta, and black for the approximated model). These lines intersect the Backbone Curves at points that correspond to the amplitude levels at which the PDFs are later compared.
Each subplot within Figure \ref{pdf_comp_Bouc} corresponds to one of the amplitude levels, comparing how well the approximated model (depicted with dotted lines) aligns with the Bouc-Wen model (depicted with solid lines). A perfect match would mean the dotted lines would overlay exactly on the solid lines, indicating the approximated model can replicate the Bouc-Wen behaviour accurately. However, discrepancies between the lines suggest deviations in the approximated model, potentially due to model form error implying that the approximated model's assumptions or structure may not fully encapsulate the actual system dynamics. Nevertheless, the amount of errors suggests that the identified model, although incorrect, can still predict the simulated measurement data with a reasonable degree of accuracy, as shown in Table \ref{table:pdf_bouc}.

\begin{table}[ht]
\centering
\caption{Comparison of distribution parameters for different amplitude levels.}
\begin{tabular}{ p{2cm} p{3cm} p{3cm}  }
\hline
Amplitude levels & Bouc-Wen, GEV($k$,$\sigma$, $\mu$) & MCMC samples, GEV($k$,$\sigma$, $\mu$) \\
\hline
Low & -0.372, 0.076, 1.521 & -0.919, 0.042, 1.610 \\
Mid & 0.292, 0.085, 1.635& 0.038, 0.042, 1.723 \\
High & -0.285, 0.114, 1.744 & 0.072,  0.044, 1.837 \\
\hline
\end{tabular}
\label{table:pdf_bouc}
\end{table}

The parameters of the GEV distribution include the shape ($k$), scale ($\sigma$), and location ($\mu$) parameters \cite{CANUDASROMO2018}. Table \ref{table:pdf_bouc}  provides a comprehensive understanding of the degree to which the MCMC sampled approximated model captures the characteristics of the experimental data. While there is some alignment in scale and location parameters, the differences in shape parameters, especially at low and high amplitudes, suggest that the approximated model may not fully represent the tail behaviour of the system as described by the Bouc-Wen model, pointing to areas where the model could be refined.\

The purpose of this analysis was to validate the approximate model by comparing its PDFs against those obtained from a known experimental model. Such a comparison is crucial for understanding the degree of accuracy of the approximate model. For Case 5, unlike the other four numerical cases, the underlying model was not previously established, which necessitates this form of validation.

\section{Validation of proposed methodology using experimental data}
A cantilever beam with two tip permanent magnets on both sides of the beam is considered\cite{Madinei}. The experimental setup aims at adjusting the resonance frequency of a bimorph piezoelectric energy harvester through the use of electromagnets. The primary objective is to demonstrate the concept of broadening energy harvesting capabilities on a larger scale \cite{hadi}.\

To validate the proposed methodology, variation in the first natural frequency (backbone curve) for different cases of airgaps and applied voltages are considered. Using the information derived from the backbone curve, an estimate of variation in system parameters (linear and non-linear stiffness) is generated using Bayesian approach by developing the stochastic model of the system.
\subsection{Experimental setup}
\begin{figure}[H]
\begin{center}
\includegraphics[width=60mm]{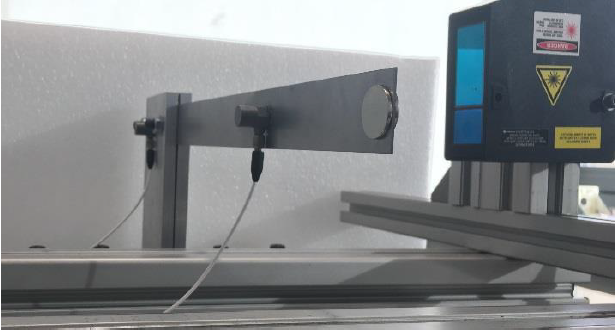}
\includegraphics[width=60mm]{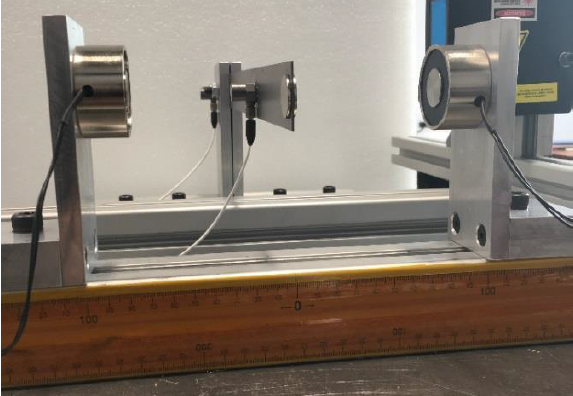}
\caption{Experimental setup of the cantilever beam with tip magnets (left) and the same beam with additional electromagnets (right)}
\label{experiment}
\end{center}
\end{figure}

The setup involves a cantilever beam equipped with two permanent magnets on each side of the tip along with electromagnets on each side (Figure \ref{experiment}). Two ICP sensors were employed to assess the acceleration at both the base and the beam, with one sensor placed at the beam's fixed end and the other situated at its midpoint.
Depending on the arrangement of magnet poles, the electromagnets can create either attractive or repulsive forces between the magnets. Through experimentation, it is observed that reducing the air gap between magnets leads to non-linear behaviour in the system. By applying voltage to the electromagnets, the resonance frequency of the beam can be fine-tuned. By utilizing different force configurations, adjusting the air gap, and varying the voltage, the lowest resonance frequency of the beam can be finely adjusted within a range of 3.8 to 9.1 Hz, demonstrating the potential for energy harvesting applications.

The experimental setup in our study, as illustrated in Figure \ref{Cantilever}, comprises a cantilever beam integrated with two tip permanent magnets on both sides under electromagnetic forces exerting either attractive or repulsive forces. Key geometrical and material properties of this setup are detailed below: the length of the beam, $L$, is 312 mm; the width of the beam, $b$, is 30 mm; the thickness of the beam, $h$, is 1.1 mm; the thickness of the tip magnets, $h_0$, is 1.5 mm; the diameter of tip magnets, $D_0$, is 20 mm; Young’s modulus of the beam, $E$, is 205.5 GPa; the density of the beam, $\rho$, is 8040 kg/m$^3$; the density of permanent magnets, $\rho_0$, is 7500 kg/m$^3$; and the viscous damping, $c_a$, is 0.455 Ns/m. The airgap is varied from 50 mm to 35 mm, while the voltage applied to the electromagnets is varied from 2V to 20V to vary the electromagnetic force between the beam and the electromagnets.\

When the air gap is reduced or the voltage applied to the electromagnets is increased, the magnetic field strength increases, resulting in a stronger interaction between the magnets. This increased interaction leads to a larger electromagnetic force, which affects the deflection of the beam. The softening non-linearity induced in the system changes the frequency response of the beam. Thus with the increase in amplitude of the vibrations, the effective stiffness of the system decreases.\

\subsection{Mathematical model}
\begin{figure}[H]
\begin{center}
\includegraphics[width=100mm]{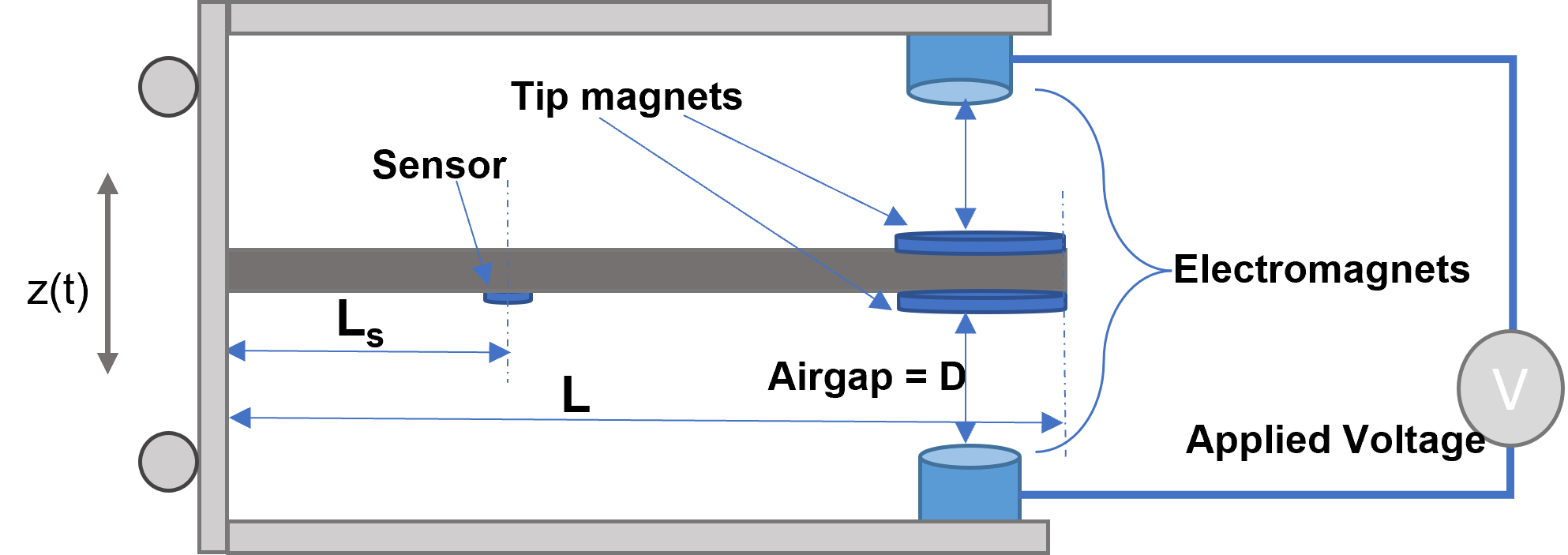}
\caption{Cantilever beam with an electromagnetic force.}
\label{Cantilever}
\end{center}
\end{figure}

This section describes the mathematical model of the cantilever beam with tip magnets under electromagnetic force (Figure.\ref{Cantilever}) that will be utilized through a Bayesian approach to generate MCMC samples. 

The simplified mathematical model used in our case is crafted to mimic the experimental testbed. The modelling approach focuses on the first mode of vibration, which is often the most significant in many structural dynamics applications. By employing the Euler-Bernoulli beam theory, which provides a classical description of bending vibrations in beams, we have concentrated on the first modal characteristics. This theoretical approach allows us to capture the essential dynamics of the system while maintaining analytical tractability.
The equation governing the transverse motion of the system can be expressed as follows:
\begin{align}
 EI \frac{\partial^4 w_{rel}(x, t)}{\partial x^4} & + \left( M_l + M_s \delta(x - L_s) \right) \frac{\partial^2 w_{rel}(x, t)}{\partial t^2} + c_a \frac{\partial w_{rel}(x, t)}{\partial t} \nonumber \\
& = -\left( M_l + M_s \delta(x - L_s) + M_t \delta(x - L) \right) \frac{\partial^2 z(t)}{\partial t^2} + F_{M} \delta(x - L)
\end{align}

Various approaches have been adopted to develop an analytical model representing the non-linear magnetic force \cite{Abdelkefi, Abdelmoula}. For the current study, to identify the non-linearity, a suitable function is presumed to represent the magnetic force ($F_M$). It's important to note that due to the symmetric arrangement of magnets at the tips, the odd terms are eliminated. Moreover, terms up to the 3rd order have been considered to enhance accuracy. Based on the experimental findings, the expression for $F_M$ is postulated as follows, where unknown parameters $\alpha_0$ and $\alpha_n$ are determined from the experimental data.
\begin{equation}
F_M=\alpha_0 w_{rel}+\alpha_n w_{rel}^3
\end{equation}

To find the dynamic response, the deflection \( w_{rel}(x, t) \) is expanded as a series in terms of the beam's eigenfunctions:
\begin{equation}
    w_{rel}(x, t) = \sum_{i=1}^{N} U_i(t)\phi_i(x)
\end{equation}

The system is subjected to the following boundary conditions:
\begin{align}
w_{\text{rel}}(0, t) &= 0, \\
\left.\frac{\partial w_{\text{rel}}}{\partial x}\right|_{x=0} &= 0, \\
\left.\frac{\partial^2 w_{\text{rel}}}{\partial x^2}\right|_{x=L} &= 0, \\
EI \left.\frac{\partial^3 w_{\text{rel}}}{\partial x^3}\right|_{x=L} &= M_t \left.\frac{\partial^2 w_{\text{rel}}}{\partial t^2}\right|_{x=L}.
\end{align}

In the above equations, \( w_{rel}(x, t) \) denotes the transverse deflection of the beam relative to its base at position \( x \) and time \( t \); \( E \) represents the Young's modulus of the beam material; \( c_a \) is the coefficient of viscous damping representing structural damping (air damping has been neglected); \( \delta(x) \) signifies the Dirac delta function; \( z(t) \) defines the base excitation function; \( M_l \) is the mass per unit length of the beam; \( M_t \) and \( M_s \) represent the mass of the tip magnets and the sensor, respectively.\

To tailor the model more closely to our experimental observations, we have reduced the model to a one-degree-of-freedom (1DOF) system. This modification is based on the realization that the system is excited at its first mode and the experimental setup does not exhibit modal interaction, which is a common occurrence in structures that are designed to operate predominantly within a single mode for the range of loading conditions considered. This simplification has led to the acquisition of a well-defined backbone curve from the experimental data, which characterizes the fundamental non-linear response of the system.\\
Applying the Galerkin method and simplifying to an approximated single DOF model \cite{Madinei}, we get:

\begin{equation}
m \ddot{U}(t) + c \dot{U}(t) + (k_m - k_0) U(t) - k_n U^3 = F_b \cos (\Omega t)
\end{equation}
Here $m$ is mass of the system of cantilever beam, $c$ is damping coefficient and $k_{m}$ represents the mechanical stiffness of the beam. They are calculated using Galerkin's approach whereas linear and non-linear stiffness, i.e. $k_{0}$ and $k_{n}$ due to the magnetic force, are derived from the experiment. $F_{b}$ is represented as the external excitation with $z_{0}$ as constant amplitude.
\begin{equation}
m = \int_0^L \left(M_l + M_s \delta(x - L_s)\right) \varphi^2(x) \, dx,
\end{equation}

\begin{equation}
k_m = E I \int_0^L \varphi(x) \varphi^{IV}(x) \, dx,
\end{equation}

\begin{equation}
c = c_a \int_0^L \varphi^2(x) \, dx
\end{equation}

\begin{equation}
F_b = z_0 \Omega^2 \left( \int_0^L (M_l + M_s \delta(x - L_s) + M_t \delta(x - L)) \phi(x) \, dx \right)
\end{equation}

\begin{equation}
\varphi(x)  =\sin \frac{\pi x}{L}
\end{equation}

For the given system $m$, $c$ and $k_{m}$ remain the same (as the variation is low) whereas stiffness parameters, i.e. $k_{o}$, $k_{n}$, vary with different airgaps and applied voltages thus changing the natural frequency.\

\begin{figure}[H]
\begin{center}
\includegraphics[width=125mm]{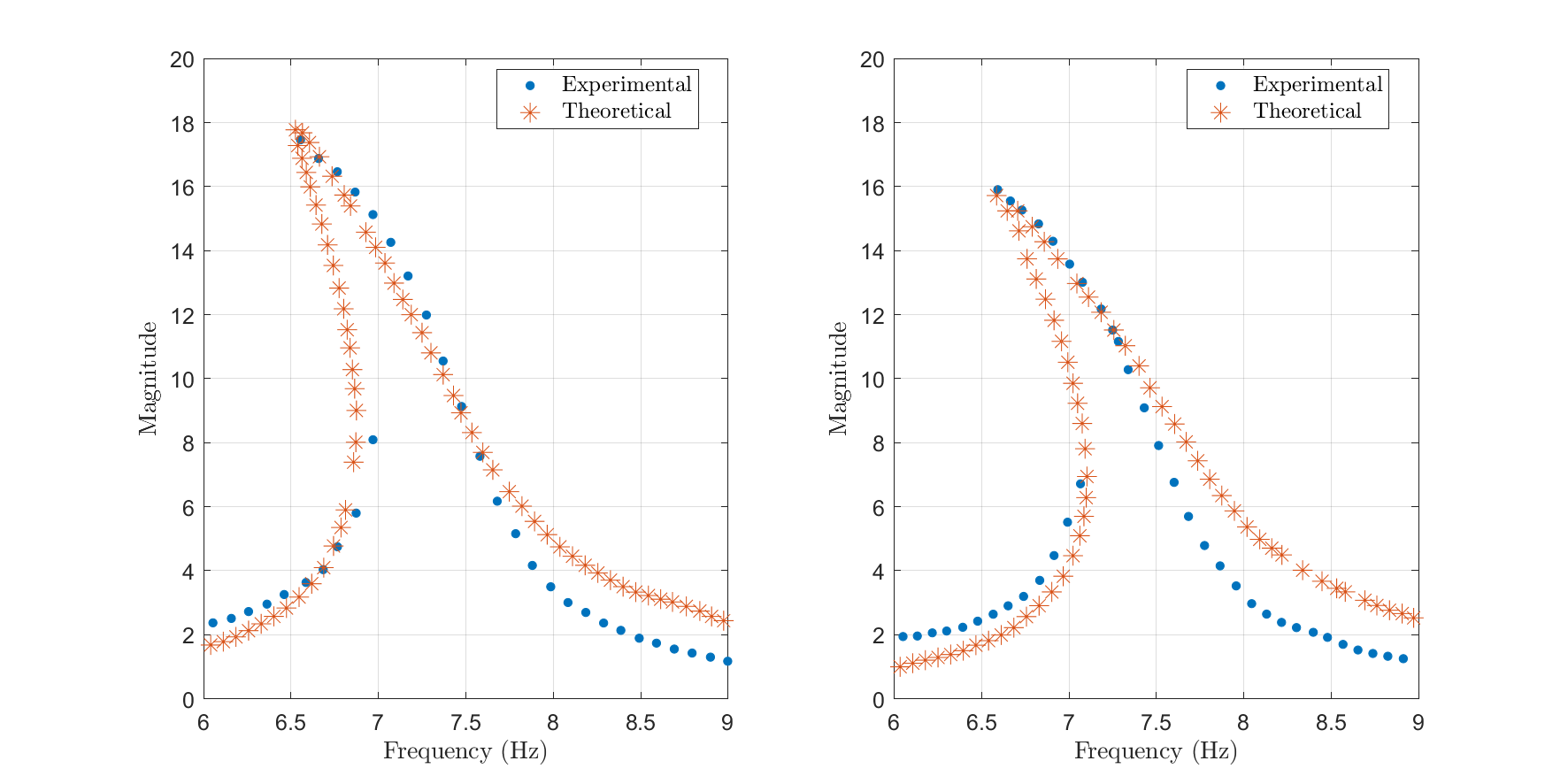}
\caption{Experimental and theoretical frequency response function (FRF) of the beam at an applied voltage of 4V (left) and 8V (right) with an airgap = 30mm.}
\label{frf}
\end{center}
\end{figure}

For airgap and applied voltage sets, the system response is measured and $k_{o}$ and $k_{n}$, which are associated with the magnetic force, are evaluated using the Matlab optimization toolbox to match the resonance frequencies under various conditions. Figure \ref{frf} demonstrates that selecting suitable $k_{o}$ and $k_{n}$ values allows the theoretical model to align with the experimental data.

Considering the low amplitude natural frequency cases and using the relation in Equation (\ref{eq:1dof}), $k_{m}$ is calculated.
\begin{equation}\label{eq:1dof}
(\textit{k}_{\textit{m}} - \textit{k}_{\textit{o}}) = \textit{m} \omega_{\textit{n}}^2
\end{equation}

Low variability in $k_{m}$ suggests that the mechanical stiffness,  remains relatively constant or experiences minimal changes compared to other parameters ($k_{o}$ and $k_{n}$) in the system. By considering the variation in $k_{m}$ as low, ($k_{m}$ - $k_{o}$) can be treated as a single parameter, simplifying the analysis and modelling of the system. 

The parameters $m = 0.03842$, $c = 0.07098$ and $k_m$ are calculated using the Galerkin approach, while $k_0$ and $k_n$ are derived from experimental data. The values of $k_0$ and $k_n$ are observed to vary with different air gaps and applied voltages, affecting the natural frequency of the system.
\begin{equation}\label{eq:dynamics}
m\ddot{x} + c\dot{x} + (\textit{k}_{\textit{m}} - \textit{k}_{\textit{o}})x - \textit{k}_{\textit{n}}x^3 = 0
\end{equation}

Using a 1DOF model (Equation (\ref{eq:dynamics})) and experimentally derived system parameters ($m$, $c$, $k_{m}$, $k_{o}$, $k_{n}$), backbone curves are generated for each set of stiffness parameters representing the effect of airgap and applied voltage on the dynamic response of the system. These backbone curves form the measurement/observation for the MCMC sampling.

\textcolor{ForestGreen}{The versatility of finite element models in representing more complex engineering structures is acknowledged. However, in this work, the effectiveness of the proposed method was demonstrated using an analytical model applied to a beam structure as a proof of concept. This approach was chosen to establish the validity of the method in a controlled, well-understood environment. The FE model is used to validate the analytical model. For simple beam structures like the one in this study, analytical models have produced results very similar to those from FE analysis. Comparison of results from the FE modelling and the analytical model are shown below:}

\begin{table}[h!]
\centering
\caption{Comparison of first three natural frequencies from FE modelling and analytical model}
\begin{tabular}{ ll }
\hline
\textbf{Method} & \textbf{Natural Frequencies (Hz)} \\
\hline
FE model  & 8.53, 54.09, 154.17 \\
Analytical model  & 8.51, 53.37, 149.02 \\
\hline
{Error (\%)} & 0.24, 1.35, 3.45 \\
\hline
\end{tabular}
\label{table:NatFreq}
\end{table}
 
\textcolor{ForestGreen}{Given this established consistency for such cases in Table \ref{table:NatFreq}, the analytical model was used due to its computational efficiency and its ability to provide clear insights into system behaviour. It is acknowledged that future work could benefit from applying the method to more complicated structures using finite element modelling, which would provide a more comprehensive validation and allow for the examination of computational costs in more detail, especially for larger-scale or more complex systems. At this stage, the focus was on demonstrating the fundamental feasibility of the approach.}

\subsection{MCMC implementation on experimental data}
MCMC sampling relies on the interplay between the observation, prior, and likelihood to infer the posterior distribution of the parameters. Through iterative updates, the Markov chain explores the parameter space by considering the prior, likelihood, and observed data, ultimately generating samples that represent the posterior distribution. This iterative sampling process allows for the quantification of uncertainty, enabling predictions and statistical inferences to be made based on the given data. Each component (observation, prior, and likelihood) play a crucial role in Bayesian inference.
\subsubsection{Observation}
The observation consists of the measured or collected data points which in this case are extracted from the system response at resonance. The observed data is crucial as it narrows down the range of plausible parameter values. By incorporating the observed data, MCMC sampling allows us to update our beliefs and generate posterior samples that are consistent with the observed evidence.
The set of parameters used for generating the measurement data, i.e. the backbone curves are defined in Table \ref{table:Exp_pars}. The initial conditions used to match the extracted backbone curves are Displacement = 0.00001 m and Velocity = 0. m/s

\begin{figure}[H]
\begin{center}
\includegraphics[scale= 0.42]{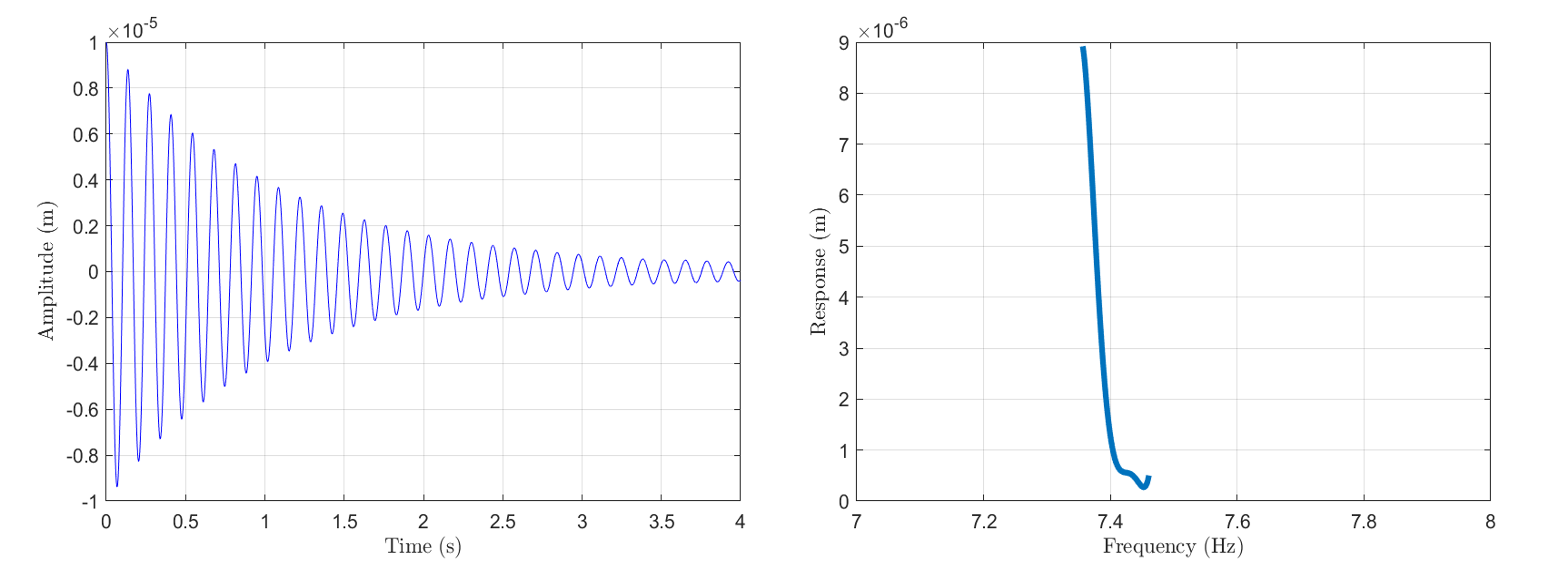}
\caption{Backbone curves from experimental response for $k_{m}-k_{0}$ = 82.6 N/m and $k_{n}$ = 9.159 \textnormal{GN/m$^3$}.}
\label{Exp_BB}
\end{center}
\end{figure}

\begin{figure}[H]
\begin{center}
\includegraphics[width=100mm]{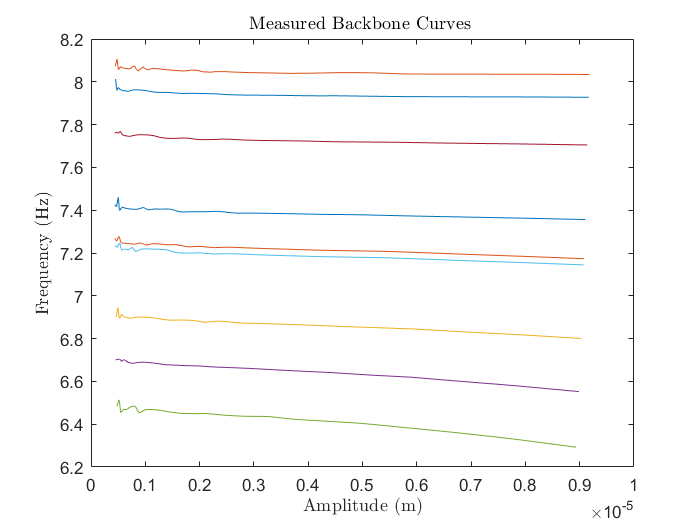}
\caption{Experimental backbone curves}
\label{bb}
\end{center}
\end{figure}

\begin{table}[ht]
\centering
\caption{Parameters used for generating backbone curves obtained by fixing base amplitude and varying airgap and applied voltage.}
\begin{tabular}{ p{2cm} p{3cm} p{3cm}  }
\hline
Samples & $k_{m} - k_{o}$ (N/m) & $k_{n}$ (\textnormal{GN/m$^3$}) \\
\hline
Sample 1 & 82.59 & 9.16 \\
Sample 2 & 79.00 & 16.30 \\
Sample 3 & 71.58 & 25.60 \\
Sample 4 & 67.28 & 38.10 \\
Sample 5 & 62.91 & 49.30 \\
Sample 6 & 78.34 & 15.86 \\
Sample 7 & 90.35 & 5.72 \\
Sample 8 & 95.33 & 0.75 \\
Sample 9 & 97.87 & 0.05 \\
\hline
\end{tabular}
\label{table:Exp_pars}
\end{table}
Figure \ref{Exp_BB} shows the extraction of the backbone curve from RDM using experimentally derived parameters. Similarly, a set of such backbone curves generated from the experimentally derived set of system parameters is shown in Figure \ref{bb}. It showcases the backbone curves derived from experimental data. These curves represent potential system responses under various conditions, highlighting the non-linear behaviour of the system. They are critical for solving the inverse problem of determining the parameters of interest and serve as essential measurements in the MCMC sampling. These backbone curves are the measurements/observations in the MCMC sampling.\


\subsubsection{Prior}
A well-informed and suitable prior can steer the posterior distribution towards parameter values that are both meaningful and plausible.  For the current analysis, a uniform distribution is used as it is considered as a non-informative prior in MCMC sampling. It assigns equal probability to all values within a specified range, making it useful when there is limited prior knowledge or a desire to remain neutral. The bounds for the uniform prior was considered by analyzing the range and distribution of the observed data and by performing sensitivity analysis on different choices of bounds.

\subsubsection{Likelihood}
\begin{figure}[H]
\begin{center}
\includegraphics[width=100mm]{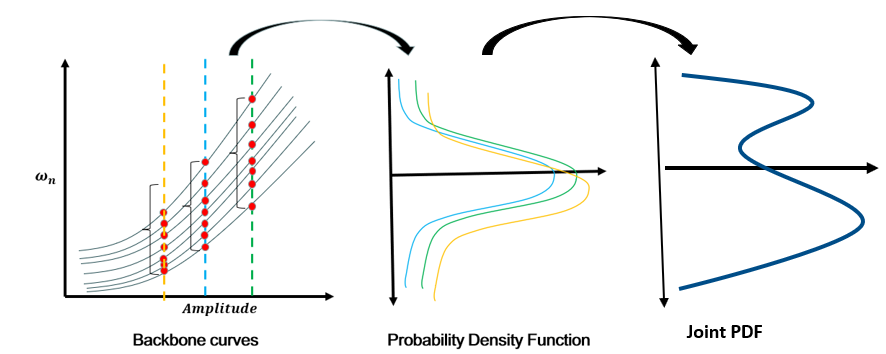}
\caption{Likelihood function. }
\label{like}
\end{center}
\end{figure}

The likelihood function expresses the probability of observing the data given the parameters of interest and quantifies the relationship between the observed data and the unknown parameters. It represents the probability of obtaining the observed data for different values of the parameters.
Figure \ref{like} provides a comprehensive illustration of how the likelihood is defined.

\subsection{Results from experimental data}
\begin{figure}[H]
\begin{center}
\includegraphics[width=120mm]{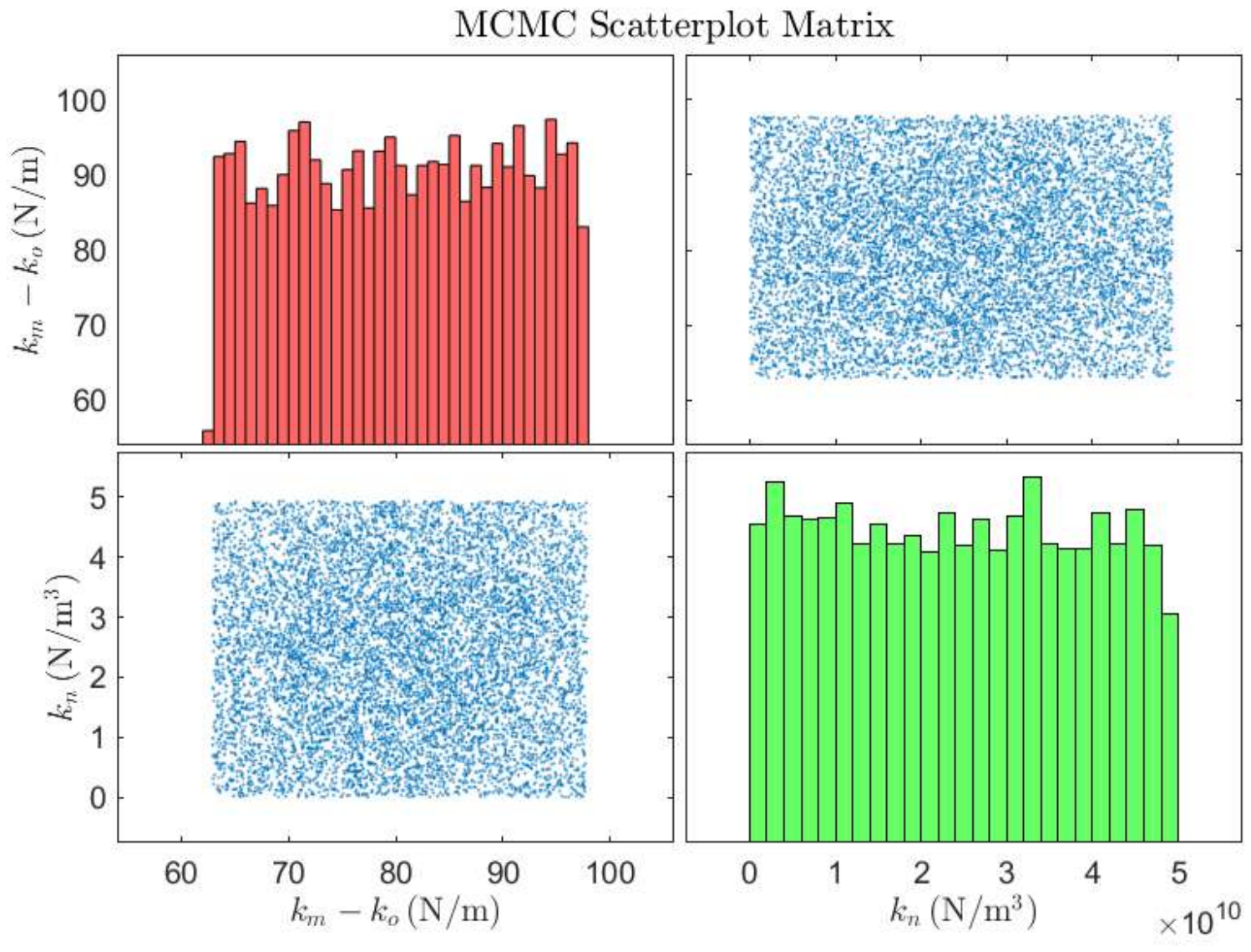}
\caption{Scatter plot Matrix of 20000 MCMC samples of $k_{m}- k_{o}$ (Red), $k_{n}$ (Green).} \label{scatter}
\end{center}
\end{figure}

After running the MCMC algorithm with the Metropolis-Hastings method, a total of $20000$ samples were obtained from the target distribution. The samples of the parameters considered are $k_{m} - k_{o}$ and $k_{n}$. Figure \ref{scatter} presents a scatter plot matrix of 20,000 MCMC samples. It provides insights into the variability and correlation of the parameters estimated through the MCMC sampling process.

In the top left, the histogram reveals the frequency of \( k_{m} - k_{o} \) values, which displays a uniform distribution, suggesting an even spread across the observed range. The top right scatterplot illustrates the relationship between \( k_{m} - k_{o} \) and \( k_{n} \), lacking any discernible trend or correlation. The bottom left offers a rotated view of the same scatterplot, essentially swapping the axes. In the bottom right, the histogram for \( k_{n} \) similarly shows a relatively uniform distribution of values. Collectively, these plots suggest that the MCMC samples adequately capture the true range of the system's parameters, as the span of the parameters \( k_{m} - k_{o} \) and \( k_{n} \) is consistent with the span of the measured data.

\textcolor{ForestGreen}{Figure \ref{scatter} represents experimental data, it was found that the experimentally derived system parameters (\( k_{m} \), \( k \), and \( k_{n} \)) also exhibited a uniform distribution. This finding aligns with the hard edges observed in the posterior samples for these parameters.  Figure \ref{20000scatterplot_duff} - \ref{20000scatterplot_bouc} correspond to known numerical cases, where backbone curves (measurement data) were intentionally generated using uniformly sampled parameters. This approach was chosen to verify our MCMC method's ability to accurately reconstruct a known target distribution. It is acknowledged that the study used a relatively modest sample size of 20,000 MCMC samples. It's possible that with a larger number of samples, the edges of the distributions might appear less pronounced. However, the clear edges primarily reflect the underlying uniform distributions rather than a limitation of the sample size.}

A convergence analysis was performed to assess the convergence of the Markov chain and ensure that a sufficient number of burn-in iterations were discarded. An acceptance level of 0.45 was obtained from 20000 MCMC samples. The Gelman-Rubin diagnostic\cite{mcmc_stats} was calculated to evaluate the mixing and convergence of the chain. The approach compares the variances between chains to the variance within each chain. It assesses the convergence of multiple chains and indicates whether the chains have reached a stationary distribution. The Gelman-Rubin diagnostic (${\hat{R}}$) was found to be 1.0004. The ${\hat{R}}$ tends to approach 1 when the chains accurately sample the target distribution. This indicates that the chains are converging to the same distribution and have reached their limiting behaviour \cite{vehtari}.

\begin{figure}[H]
\begin{center}
\includegraphics[width=100mm]{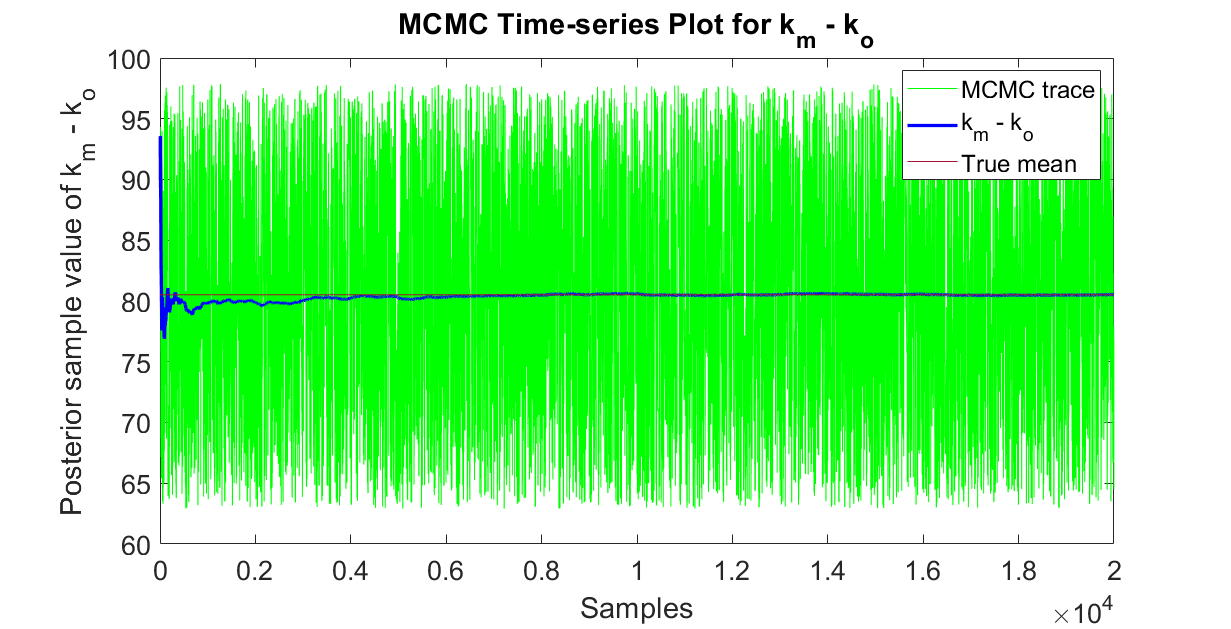}
\includegraphics[width=100mm]{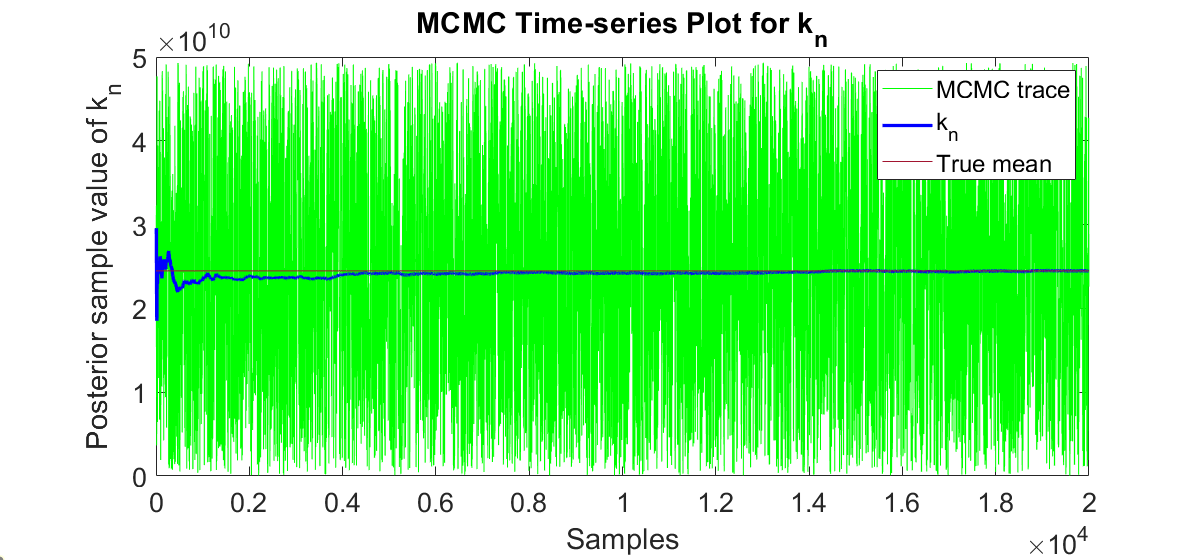}
\caption{Trace plot of MCMC samples.}
\label{trace}
\end{center}
\end{figure}

The convergence of the Markov chain was first assessed by examining trace plots and calculating convergence diagnostics such as the Gelman-Rubin statistic. The trace plot for each parameter shown in Figure \ref{trace} illustrates the values that the parameter assumed throughout the duration of the chain's execution. As depicted in the trace plot, following an initial burn-in period, the \( k_{m} - k_{o} \) and \( k_{n} \) sample values stabilize and oscillate around the true mean, implying that the MCMC chain has reached a stationary distribution. This behaviour suggests the samples post-burn-in are suitable for statistical analysis, with the high density of the trace indicating a robust exploration of the parameter space around the true mean.

Summary statistics such as posterior means, standard deviations, and credible intervals for the model parameters were computed based on the MCMC samples (Table\ref{table:summary}). These statistics provided insights into the estimated values and the associated uncertainty. The posterior mean can be interpreted as the most likely or expected value given the observed data and prior information the standard deviation measures the variability or spread of the posterior distribution around the posterior mean.\

Credible intervals provide a range of plausible values for the parameters based on the posterior distribution. They represent the uncertainty in the parameter estimation and give a measure of the parameter's precision. Here, the 95\% confidence level is taken for calculating the credible interval allowing for the quantification of uncertainty and providing a range within which the true parameter value is likely to fall.
\begin{table}[ht]
\centering
\caption{Summary statistics for the model parameters from MCMC samples.}
\begin{tabular}{ p{1.15cm} p{1.25cm} p{1.95cm} p{2.65cm}  }
\hline
Parameter & Mean & Standard Deviation & 95\% CI\\
\hline
$k_{m} - k_{o}$ (N/m)  &   $80.5$  & $10.1$ & $63.7$ - $96.9	$  \\
$k_{n}$ \textnormal{GN/m$^3$} &  $25 $ & $143.7 $ & $1.3$ - $48$ \\
\hline
\end{tabular}
\label{table:summary}
\end{table}

\begin{figure}[H]
\centering
\includegraphics[width= 60mm]{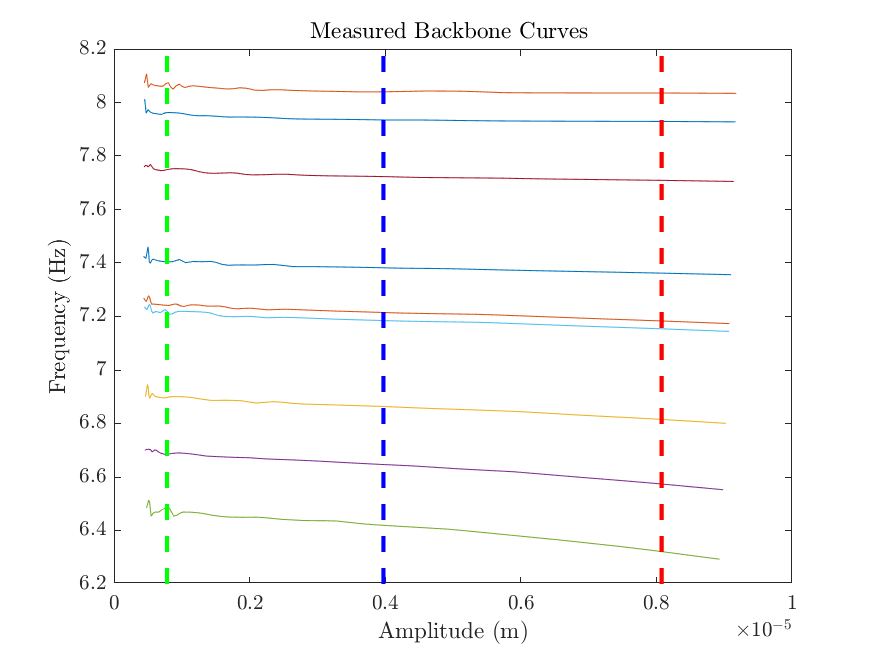}
\includegraphics[width= 60mm]{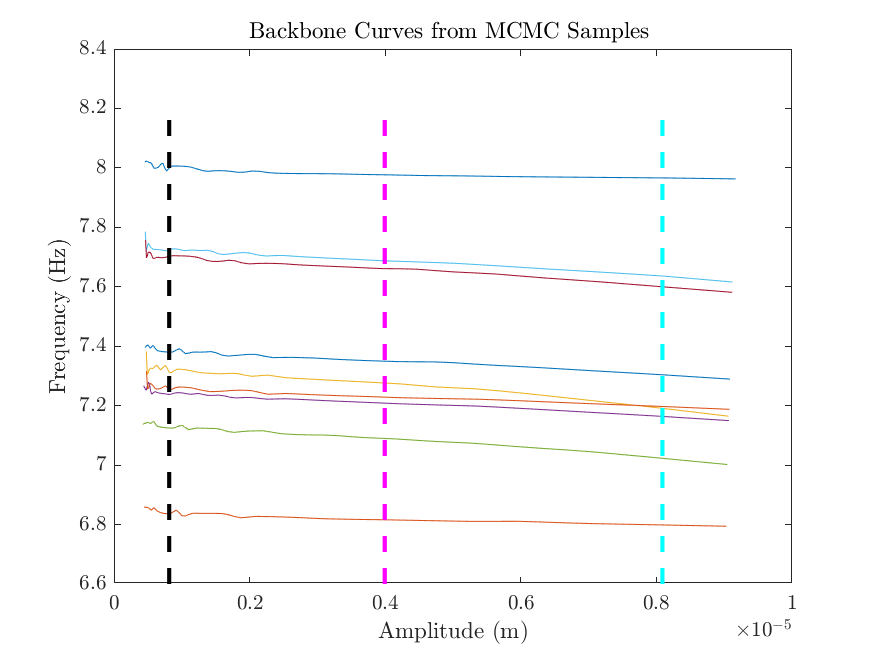}
\caption{Measured and MCMC sampled backbone curves}
\label{pdf_bb_curve_exp}
\end{figure}

\begin{figure}[H]
\centering
\includegraphics[width= 130mm]{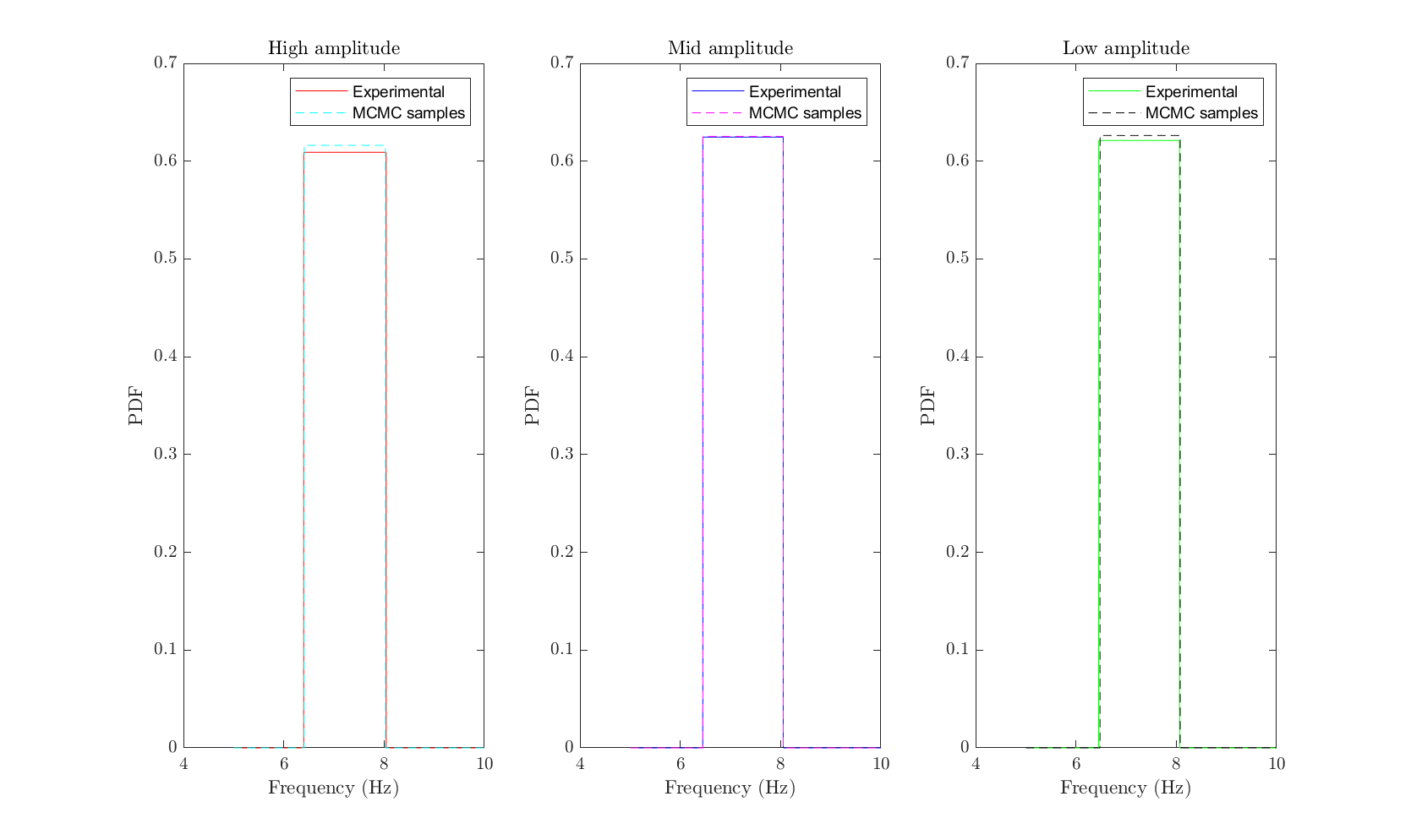}
\caption{PDF Comparison of MCMC results for experimental case}
\label{pdf_comp_canti}
\end{figure}

Furthermore, sensitivity analyses were conducted by varying the priors or model assumptions to evaluate the robustness of the results. Influential parameters were identified, and the impact of different prior specifications on the posterior distributions was assessed through sensitivity analyses. The MCMC sampling results and sensitivity analyses contribute to the understanding of the model's performance, parameter estimates, and the overall uncertainty in the analysis.
A comprehensive comparison was made between the experimental data and the results obtained from MCMC sampling at different amplitude levels. Figure \ref{pdf_bb_curve_exp} compares the measured Backbone Curves with those obtained from MCMC samples for the given model. The coloured vertical dashed lines intersecting both sets of Backbone Curves represent the amplitudes at which the detailed PDF comparison is performed. The alignment (or lack thereof) between these curves at these amplitudes serves as an indicator of the accuracy of the MCMC model in replicating the experimental observations. Figure \ref{pdf_comp_canti} presents the PDFs for the high, mid, and low amplitude levels, comparing the experimental case against the MCMC samples. The PDFs reflect the distribution of frequency responses of the system at each amplitude level. The closeness of the PDFs derived from the experimental data and the MCMC samples would suggest how well the MCMC model is able to capture the behaviour of the actual system.

Table \ref{table:pdf_exp} provides a numerical comparison of the distribution parameters for different amplitude levels, specifically detailing the parameters of the Uniform distribution used to model the experimental and MCMC-sampled data. Notably, the table indicates a high degree of similarity between the parameters of the Uniform distribution for the experimental data and the MCMC samples, with only slight variations. This suggests that the MCMC model has a good degree of accuracy in terms of the central tendency and the spread of the data, which is particularly important when models are used for predictive purposes in simulations or for understanding system behaviour under various operating conditions.

\begin{table}[ht]
\centering
\caption{Comparison of distribution parameters for different amplitude levels.}
\begin{tabular}{ p{2cm} p{3cm} p{3cm}  }
\hline
Amplitude levels & Experimental, Uniform($a$,$b$) & MCMC samples, Uniform($a$,$b$) \\
\hline
Low & 6.398, 8.040 & 6.405, 8.028 \\
Mid & 6.450, 8.051 & 6.452, 8.051 \\
High & 6.454, 8.064 & 6.481, 8.078 \\
\hline
\end{tabular}
\label{table:pdf_exp}
\end{table}

\section{Conclusion}
This study introduces a method for stochastic nonlinear model updating, employing a novel likelihood function and utilizing Bayesian inference. We use the statistics of measured backbone curves to construct these likelihood functions. To extract the backbone curves, which are then used as inputs for subsequent Markov Chain Monte Carlo (MCMC) sampling, we employed the resonant decay method. This methodology is validated both numerically and experimentally with data from a cantilever beam system fitted with permanent magnets and electromagnets. By adjusting the airgap and applied voltage,  the system's resonance frequency was changed. The MCMC sampling process was then employed to investigate the posterior distributions of the system parameters. The efficiency of the MCMC algorithm is demonstrated through trace plots and scatterplots, showcasing its proficiency in accurate sampling from the posterior distribution.

The uncertainty in the parameter estimates is effectively quantified using summary statistics, including posterior means, standard deviations, and 95\% credible intervals. The analysis validates the likelihood function's ability to capture parameter variations and confirms the reliability of MCMC sampling for obtaining accurate parameter estimates and understanding the associated uncertainty. This study offers valuable insights into the behaviour of the analyzed system and its estimated parameters, enhancing our understanding of similar systems.

This paper demonstrated that the newly proposed likelihood function is suitable for stochastic nonlinear model updating.  This makes it relevant to a broader range of dynamic systems and deepens our comprehension of their stochastic nature. This advancement underscores the methodology's adaptability and practicality in analyzing and modelling complex dynamical systems characterized by uncertainty. Furthermore, this framework can be adapted to different nonlinear models for quantifying uncertainty in parameters, thereby aiding in the development of stochastic models.

\textcolor{ForestGreen}{While the method introduced a novel approach using backbone curves in MCMC analysis for stochastic model updating, the importance of comparative studies is also recognised. Future work could explore ways to benchmark the method against existing MCMC-integrated Bayesian techniques, despite the fundamental differences in approach. This comparison could further highlight the strengths and unique contributions of the backbone curve-based methodology in the field of stochastic nonlinear model updating.}

\section{Acknowledgement}
This research work has been partly funded by the RCUK Energy Programme [grant number EP/T012250/1]. The author also acknowledges the funding from the Engineering Physical Science Research Council (EPSRC) through a program grant EP/R006768/1. The views and options expressed herein do not necessarily reflect those of UKAEA. The support of the Supercomputing Wales project, which is part-funded by the European Regional Development Fund (ERDF) via the Welsh Government is also acknowledged.

\bibliographystyle{ieeetr}
\bibliography{Bibliography_stochastic}

\section{Statements \& Declarations}
\subsection{Funding}
This research work has received partial funding from the RCUK Energy Programme under grant number EP/T012250/1. The author also expresses gratitude for the financial support provided by the Engineering Physical Science Research Council (EPSRC) through program grant EP/R006768/1.
\subsection{Competing interest}
The authors confirm that they do not have any identifiable financial conflicts of interest or personal affiliations that might have seemed to exert an influence on the research presented in this paper.
\subsection{Author contribution}
\textbf{Pushpa Pandey}: Data curation, Formal analysis, Investigation, Methodology, Software, Validation, Visualization, Writing – original draft, Writing – review \& editing. \textbf{Prof Hamed Haddad Khodaparast} and \textbf{Prof Michael Ian Friswell}: Conceptualization, Data curation, Funding acquisition, Investigation, Project administration, Supervision, Visualization, Writing – review \& editing. \textbf{Dr Tanmoy Chatterjee}: Supervision, Writing – review \& editing. \textbf{Hadi Madinei}: Data curation, Formal analysis, Investigation. \textbf{Tom Deighan}: Supervision.

\section{Data availability}
The datasets analysed during the current study are not publicly available but can be made available on request.

\section{Appendix}
\subsection{Results from MCMC samples with ODE Solver}
\label{MCMC_results_ODE}
\begin{figure}[H]
\includegraphics[width=0.5\textwidth]{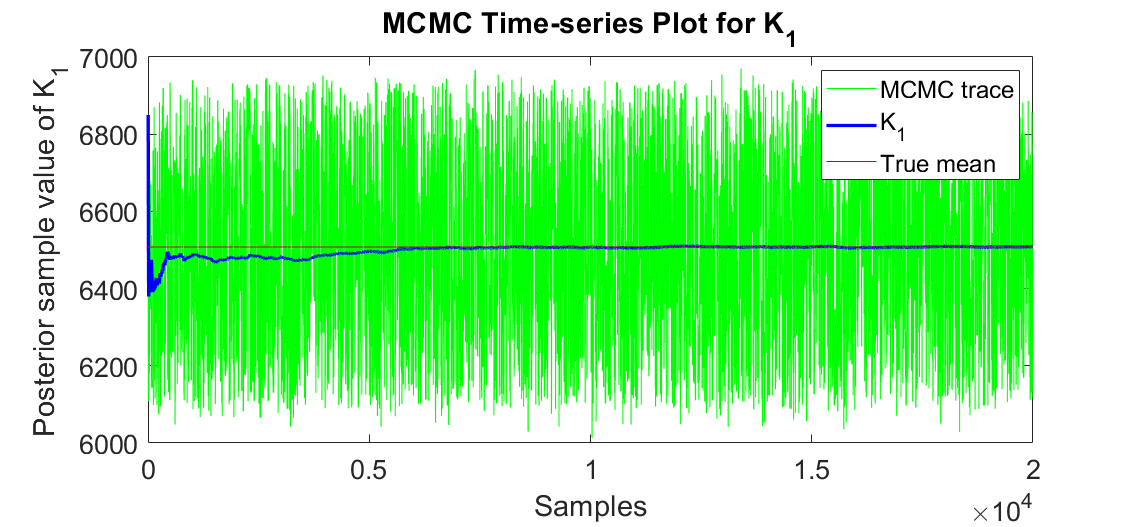}
\includegraphics[width=0.5\textwidth]{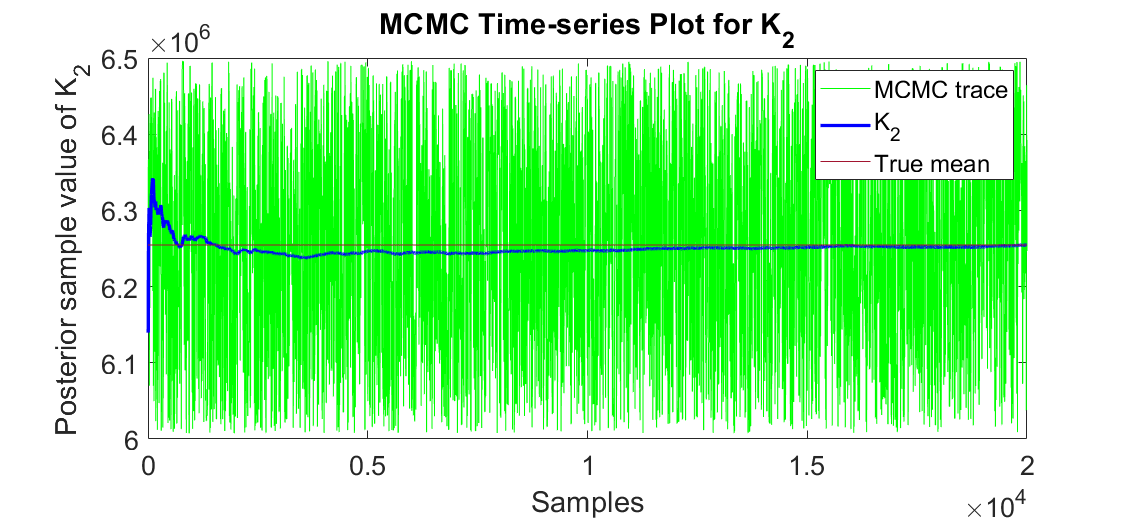}
\begin{center}
\includegraphics[width=0.5\textwidth]{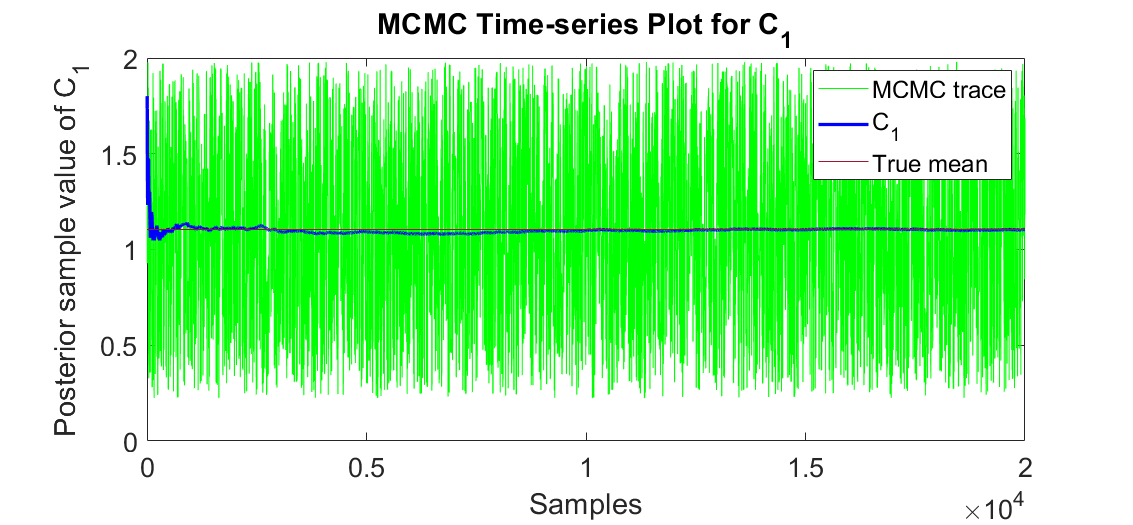}
\end{center}
\caption{Trace plot for $k_{1}$, $k_{2}$, and $c_{1}$ with 20000 MCMC samples for Case 1.}
\label{traceplot20000}
\end{figure}

\begin{figure}[H]
\includegraphics[width=0.5\textwidth]{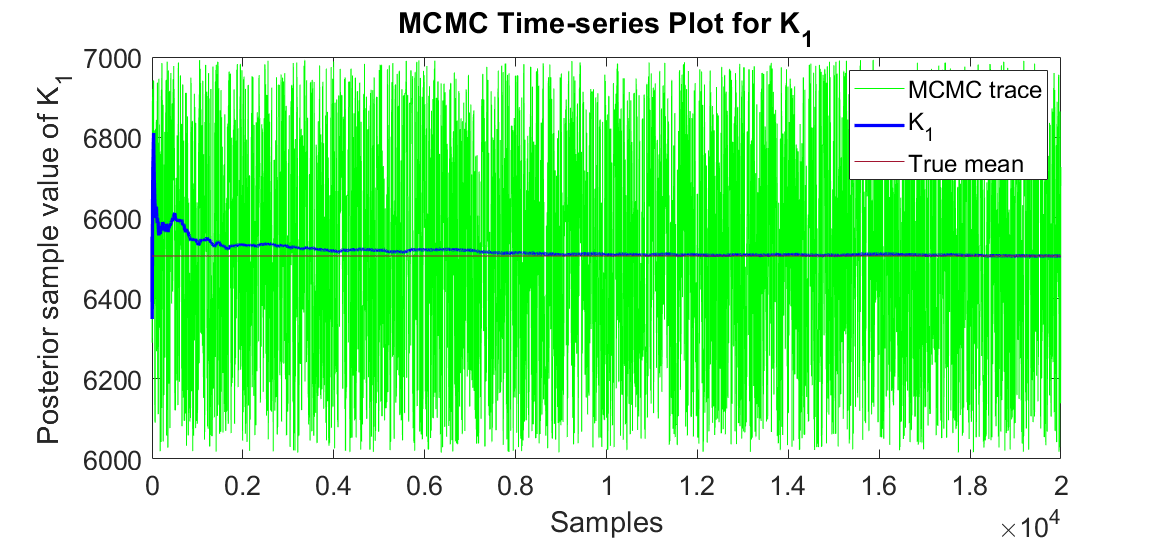}
\includegraphics[width=0.5\textwidth]{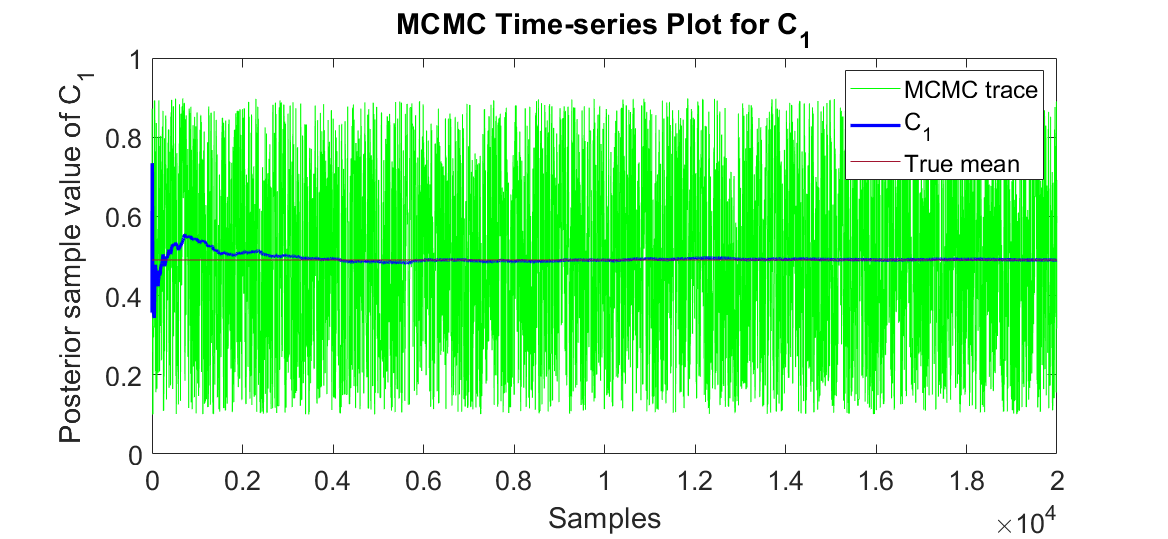}
\begin{center}
\includegraphics[width=0.5\textwidth]{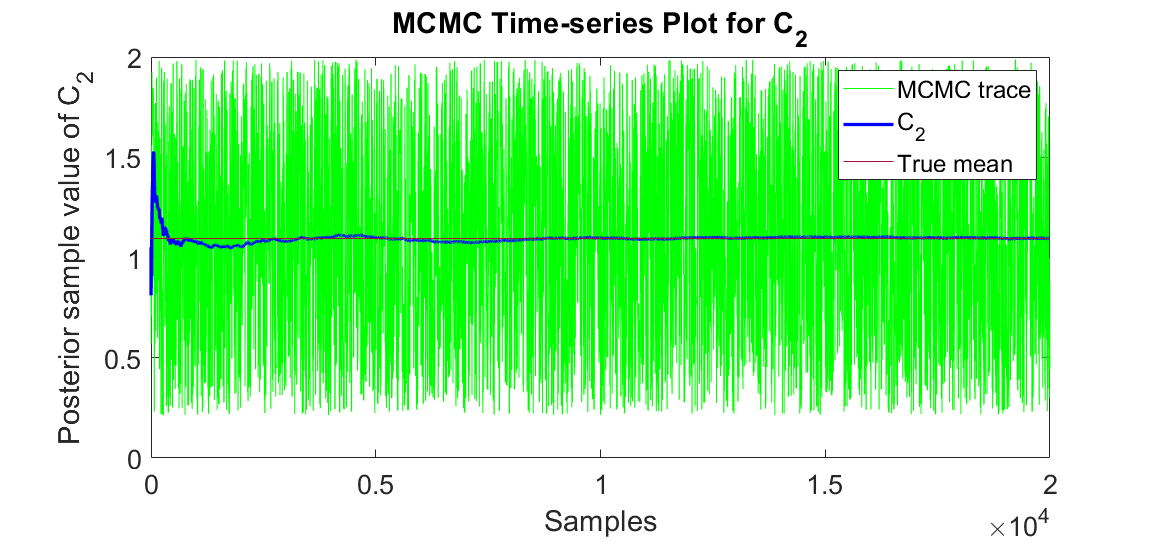}
\end{center}
\caption{Trace plot for $k_{1}$, $c_{1}$ and $c_{2}$ with 20000 MCMC samples for Case 2.}
\label{traceplot_dry}
\end{figure}

\begin{figure}[H]
\includegraphics[width=0.5\textwidth]{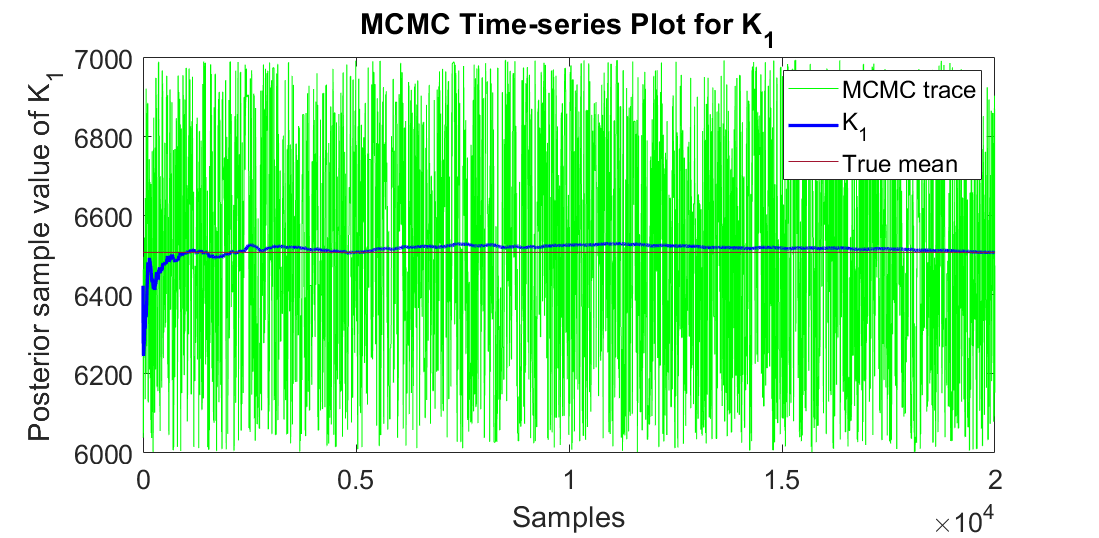}
\includegraphics[width=0.5\textwidth]{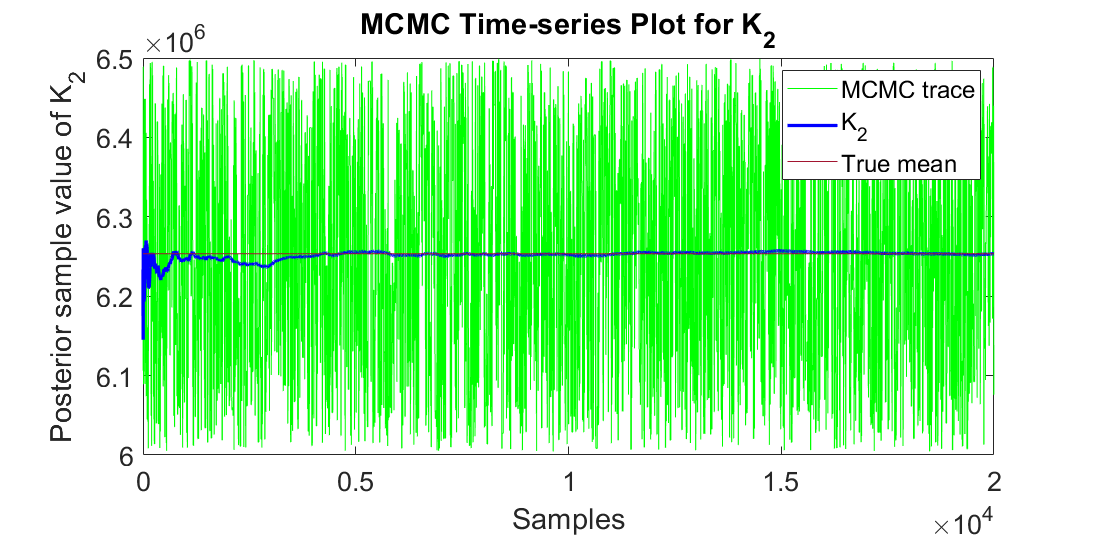}
\includegraphics[width=0.5\textwidth]{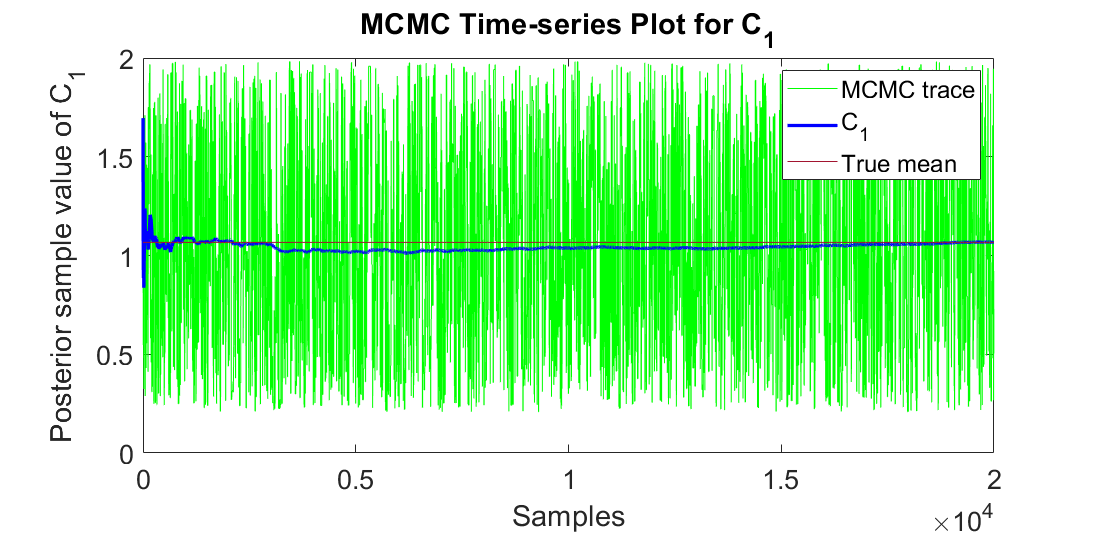}
\includegraphics[width=0.5\textwidth]{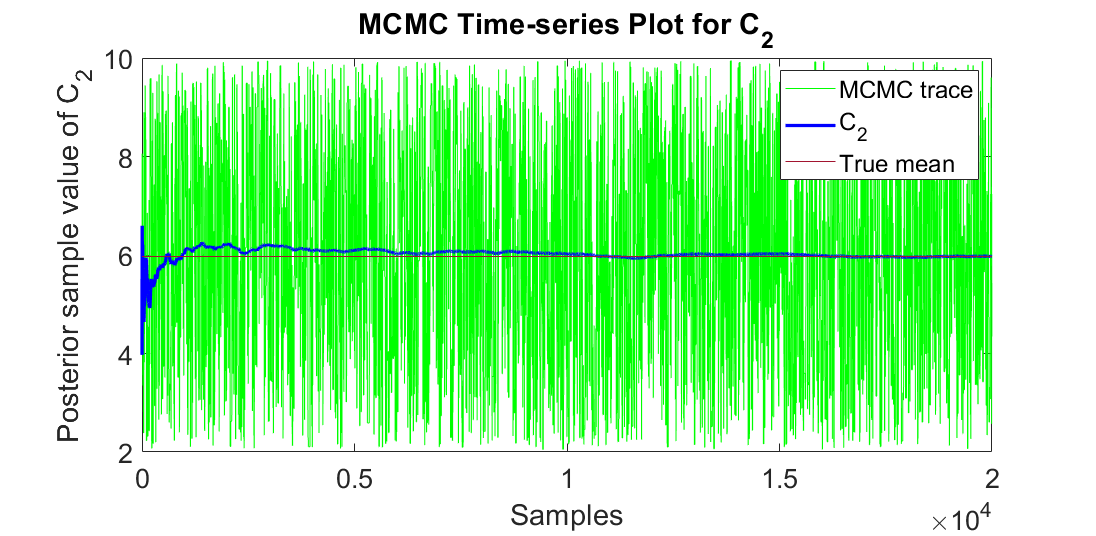}
\caption{Trace plot for $k_{1}$, $k_{2}$, $c_{1}$ and $c_{2}$ with 20000 MCMC samples for Case 3.} \label{traceplot20000quad}
\end{figure}

\begin{figure}[H]
\includegraphics[width=0.5\textwidth]{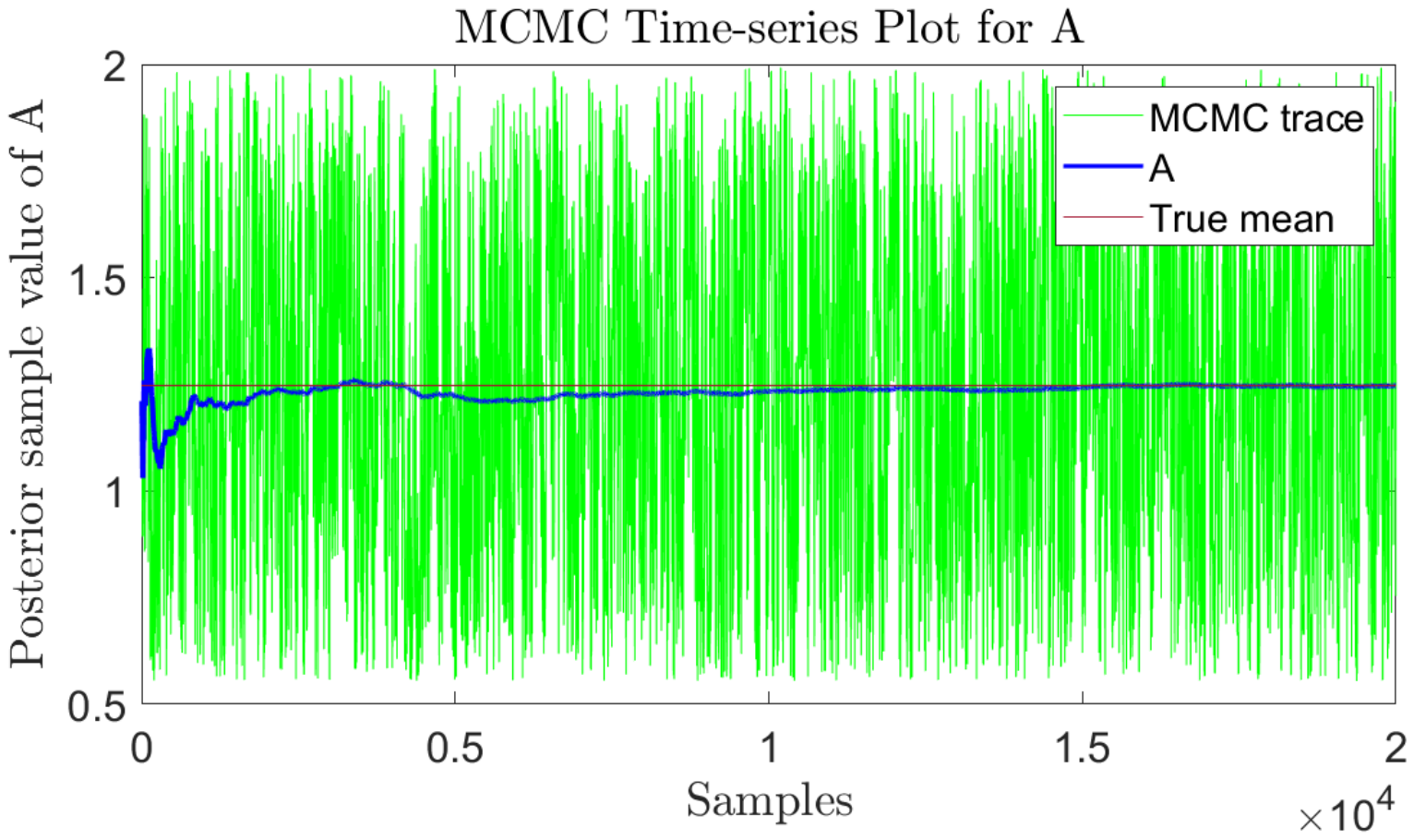}
\includegraphics[width=0.5\textwidth]{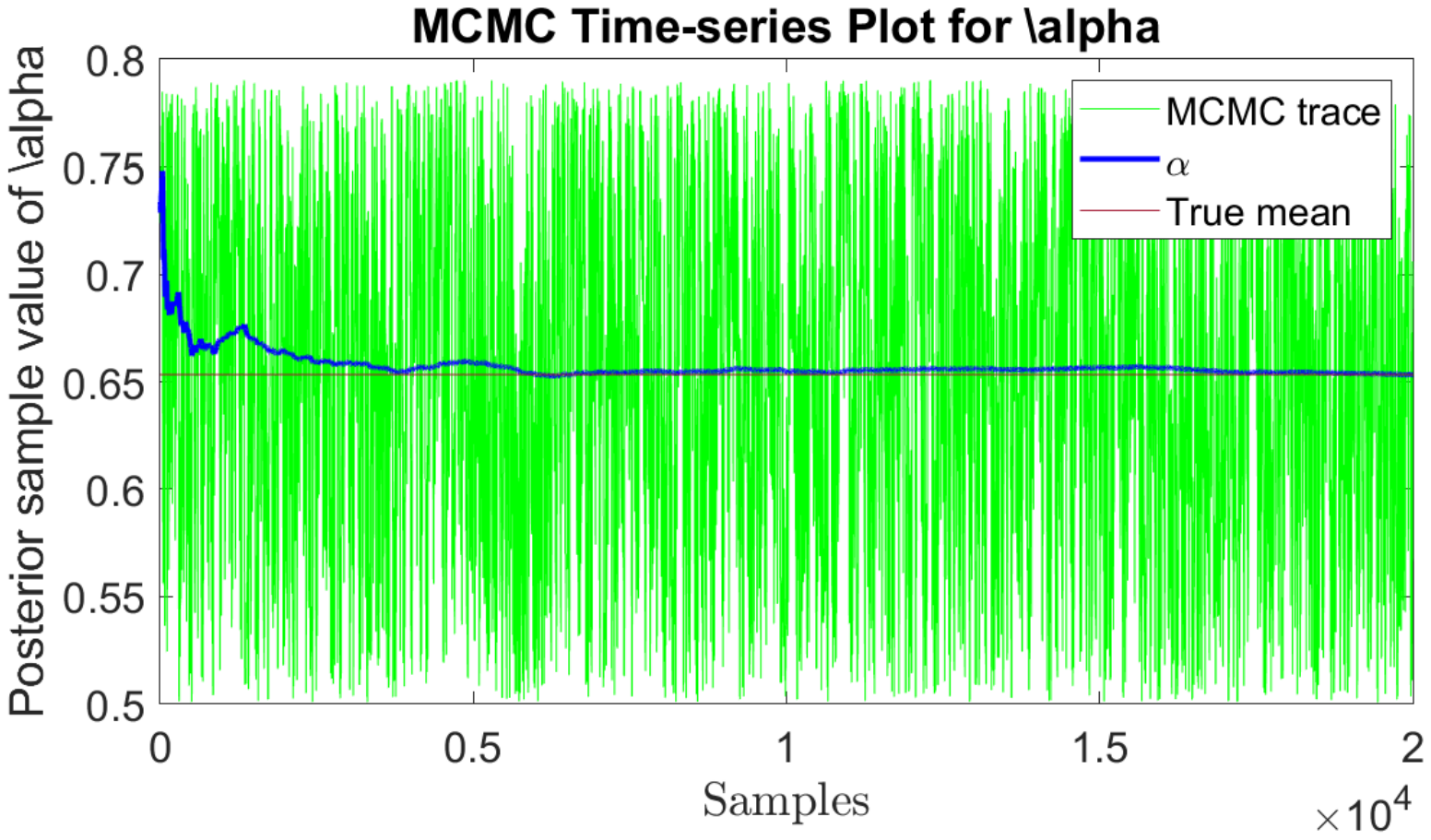}
\includegraphics[width=0.5\textwidth]{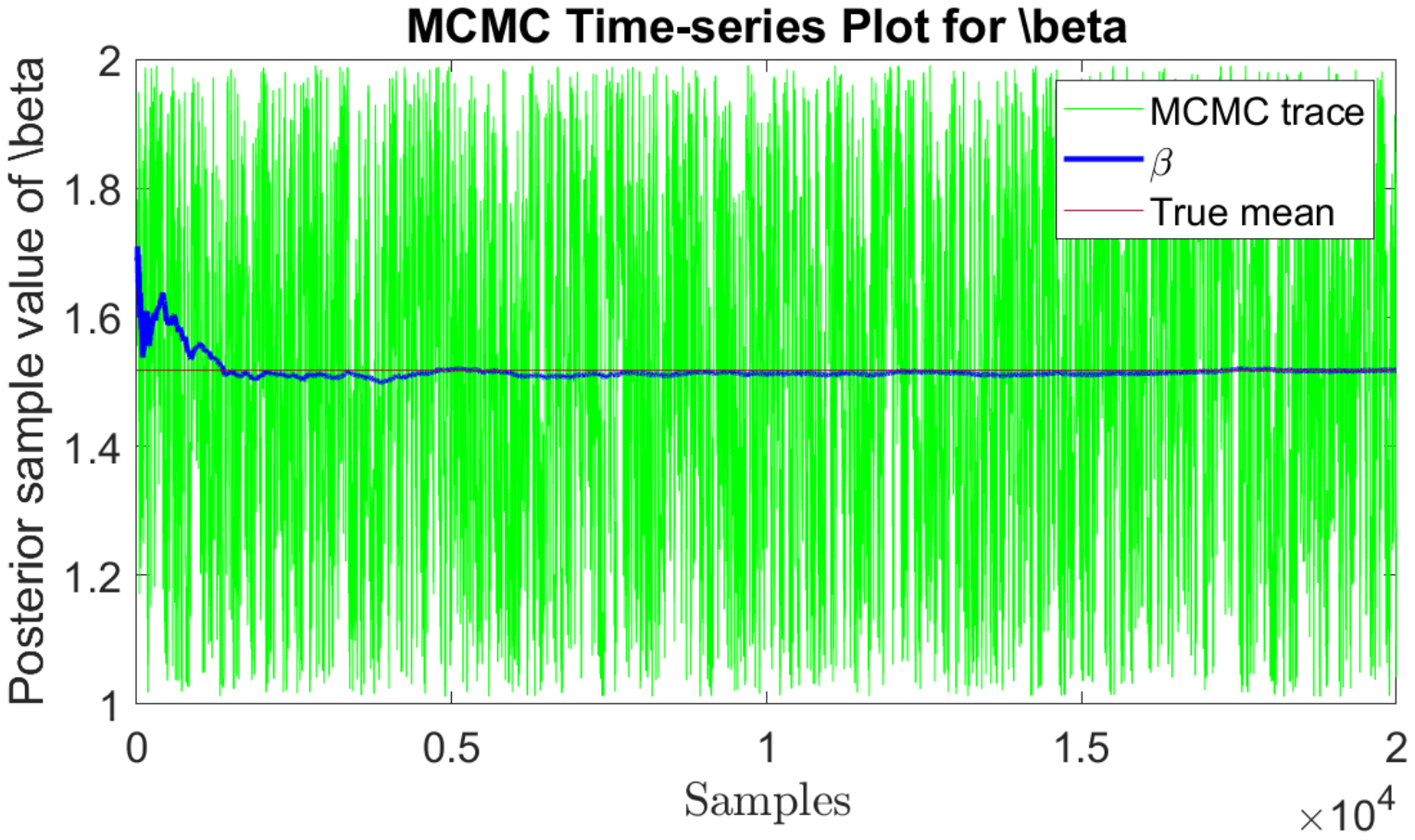}
\includegraphics[width=0.5\textwidth]{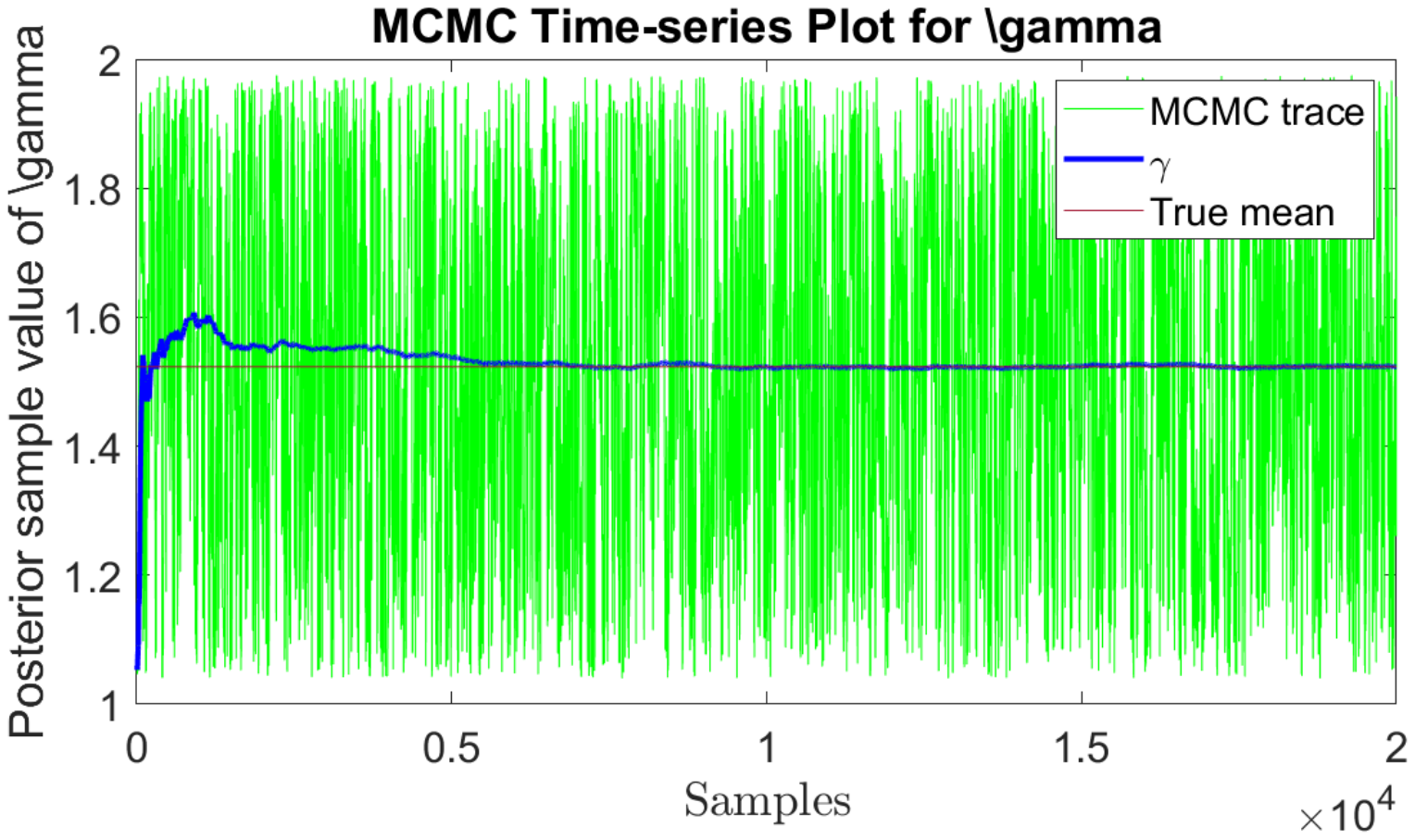}
\caption{Trace plot for (\( A\), \(\alpha \), \( \beta\) and \(\gamma\) with 20000 MCMC samples for Case 4.} \label{traceplot_bouc}
\end{figure}

\begin{figure}[H]
\includegraphics[width=0.5\textwidth]{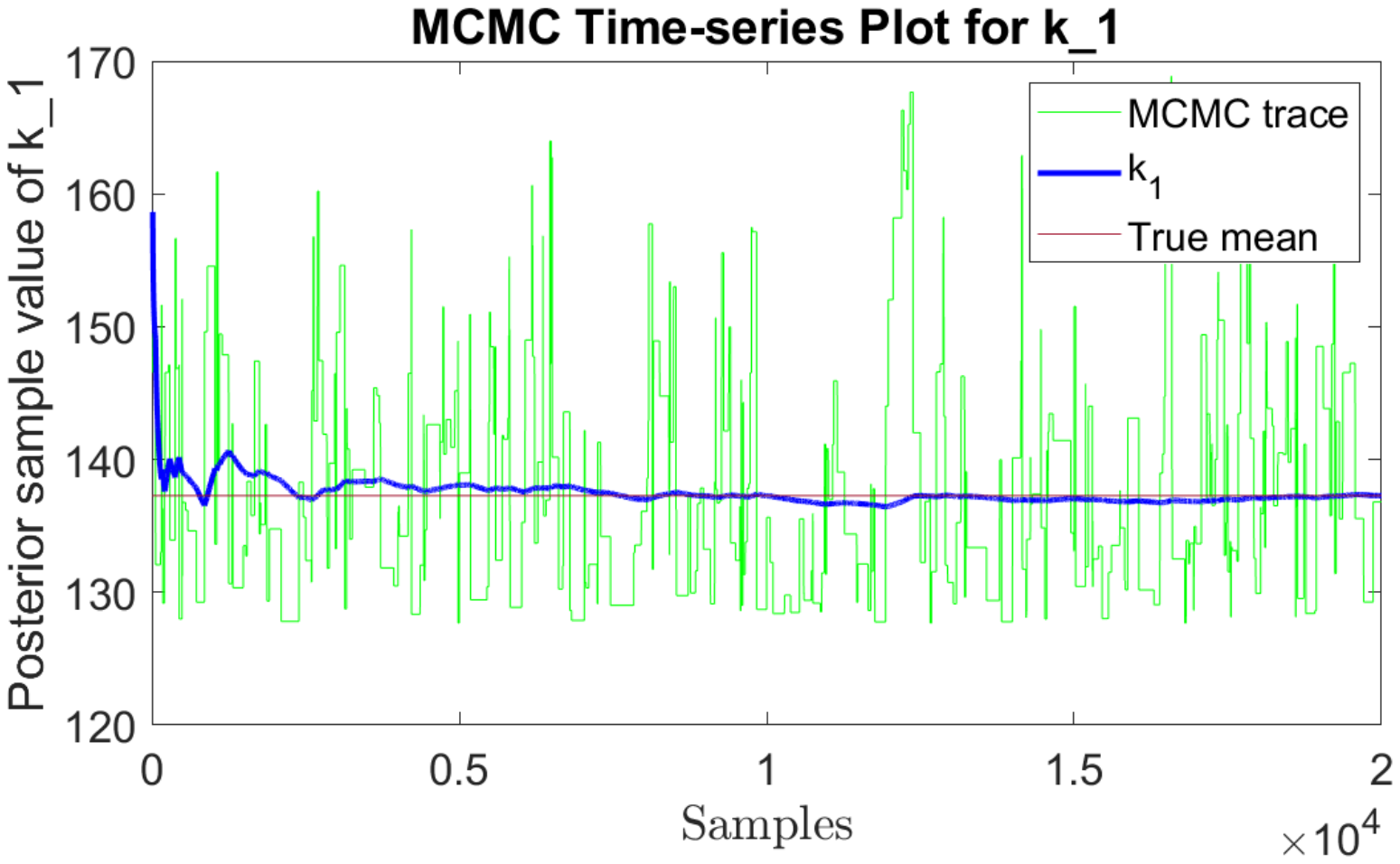}
\includegraphics[width=0.5\textwidth]{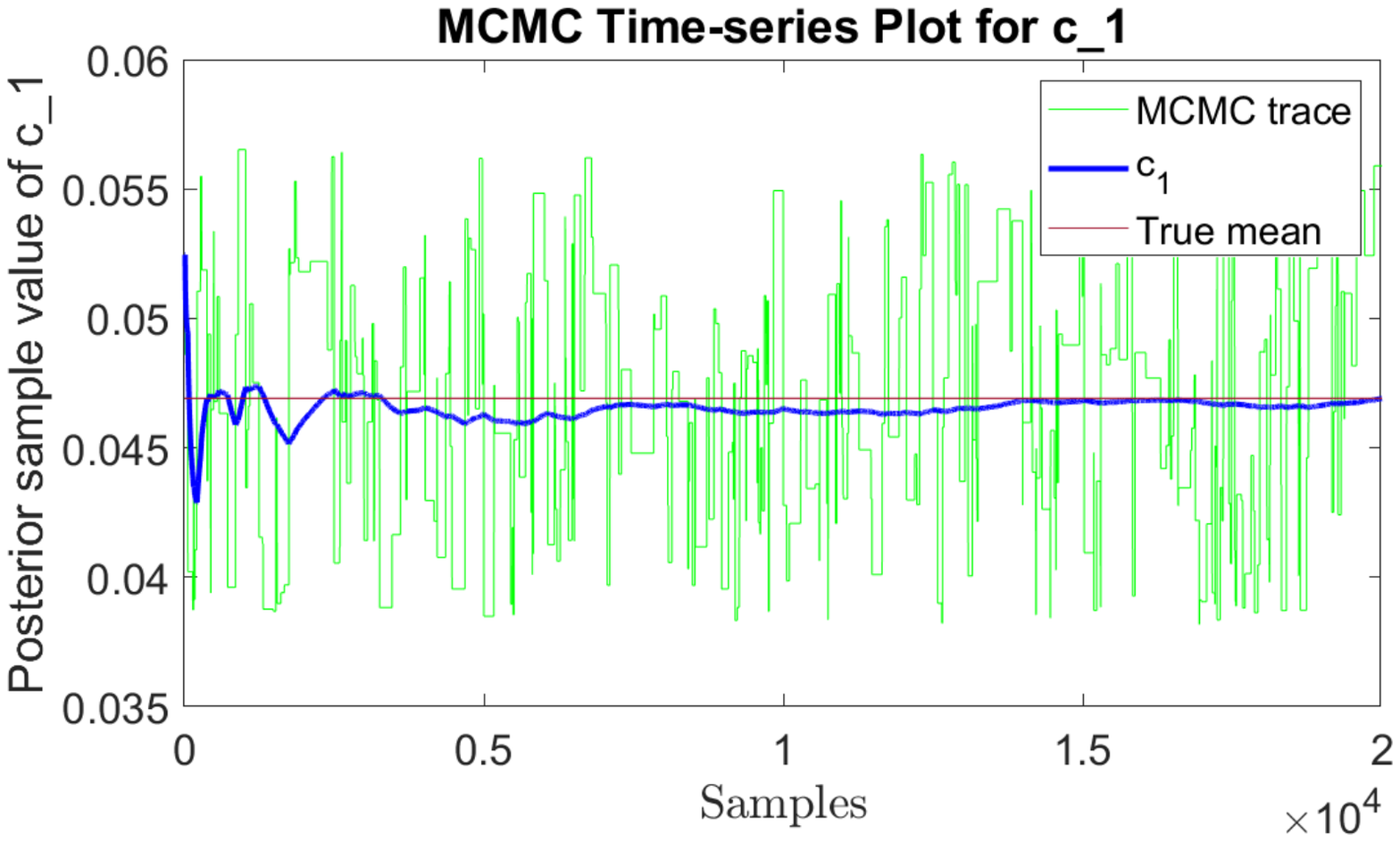}
\includegraphics[width=0.5\textwidth]{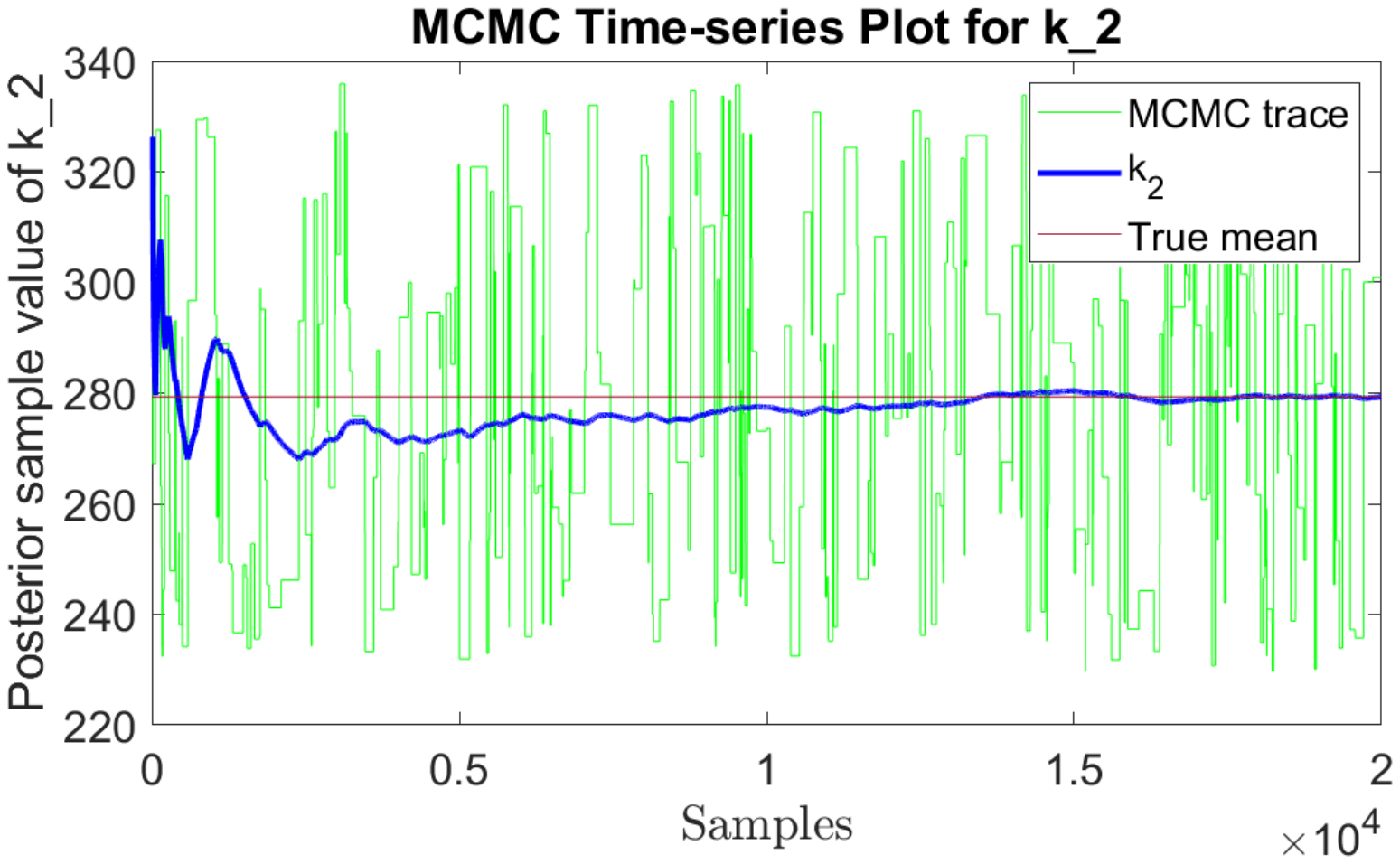}
\includegraphics[width=0.5\textwidth]{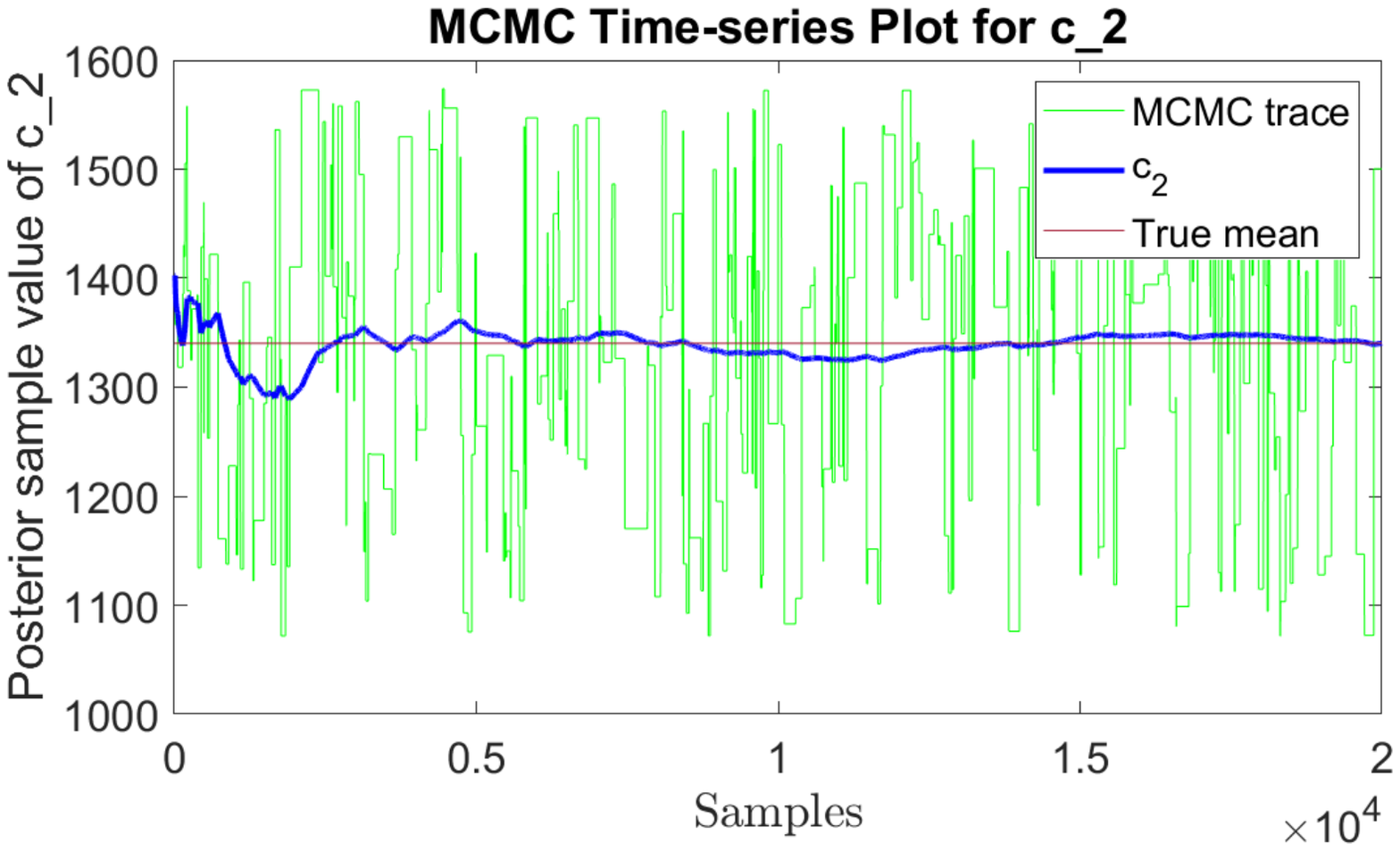}
\includegraphics[width=0.5\textwidth]{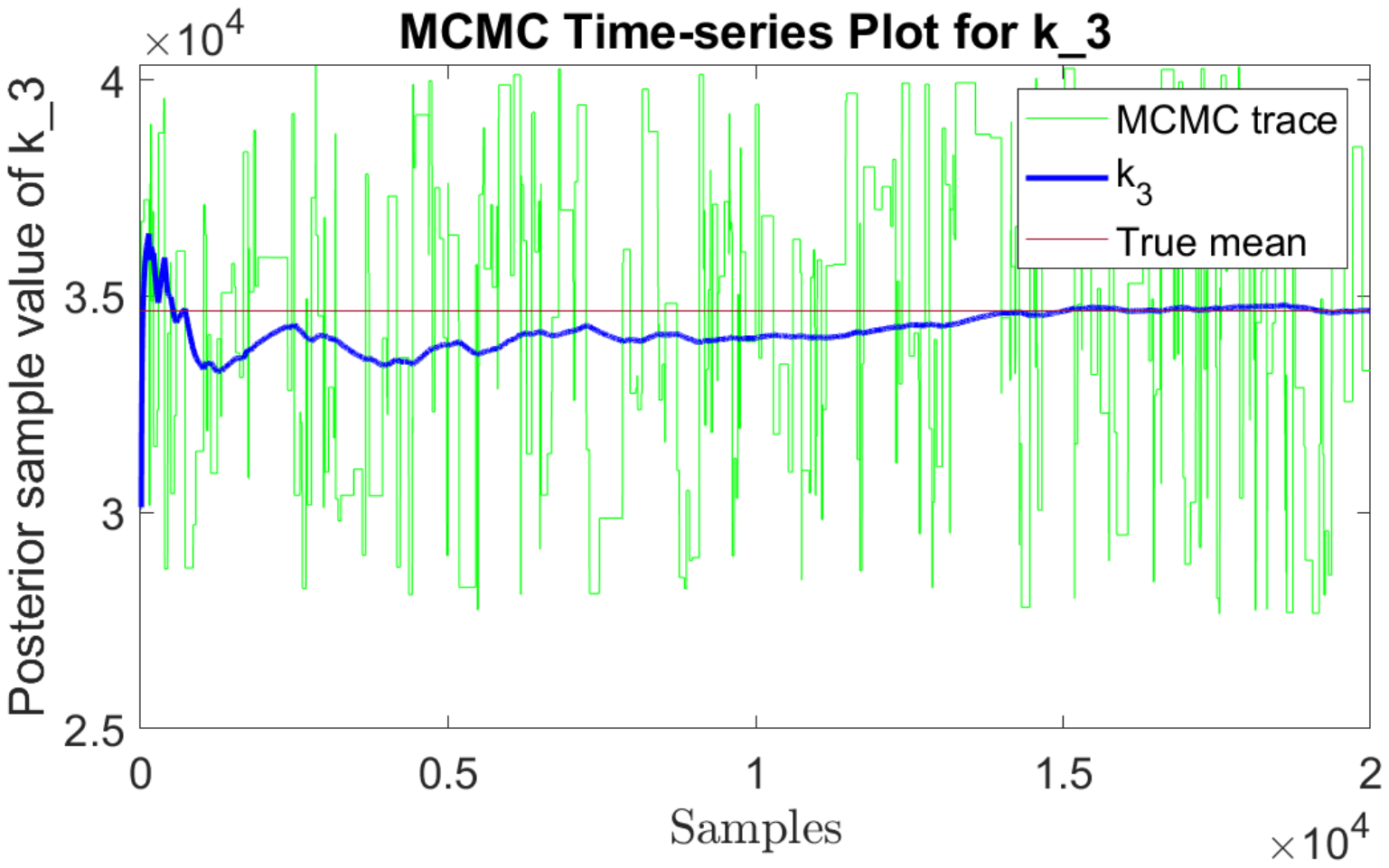}
\includegraphics[width=0.5\textwidth]{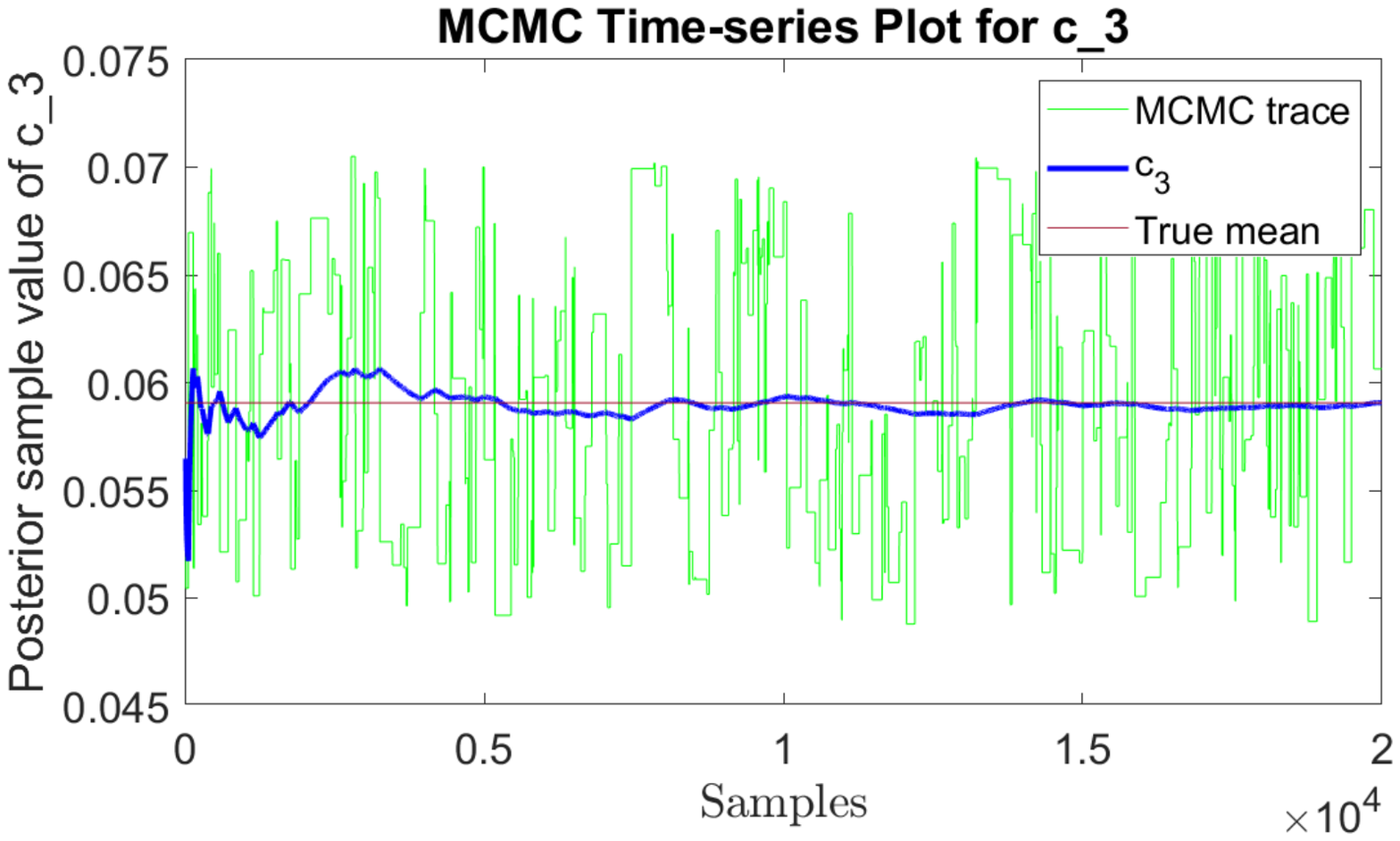}
\caption{Trace plot for $k_{1}$, $c_{1}$, $k_{2}$, $c_{2}$,  $k_{3}$ and $c_{3}$ with 20000 MCMC samples for Case 5.}
\label{traceplot20000approx}
\end{figure}

\end{document}